\documentclass[preprint,12pt,authoryear]{elsarticle}

\usepackage{amssymb}
\usepackage{rotating}
\usepackage{amsmath}
\usepackage{url}
\usepackage{float}
\usepackage{hyperref} 
\usepackage[a4paper, total={7in, 10in}]{geometry}
\RequirePackage{graphicx,caption}
\graphicspath{{./images}} 
\RequirePackage[authoryear]{natbib}
\usepackage{subcaption} 

\journal{Applied Energy}

\begin{document}

\begin{frontmatter}



\title{A sub-hourly spatio-temporal statistical model for solar irradiance in Ireland using open-source data} 


\author[label1]{Maeve Upton} 
\author[label1]{Eamonn Organ}
\author[label2]{Amanda Lenzi}
\author[label1]{James Sweeney}
\affiliation[label1]{organization={MACSI, Department of Mathematics and Statistics, University of Limerick},
            country={Ireland}}
\affiliation[label2]{organization={MISS,  University of Edinburgh},
            country={UK}}

\begin{abstract}
Accurate estimation of solar irradiance is essential for reliable modelling of solar photovoltaic (PV) power production. In Ireland’s highly variable maritime climate, where ground-based measurement stations are sparsely distributed, selecting an appropriate solar irradiance dataset presents a significant challenge. This study introduces a novel Bayesian spatio-temporal modelling framework for predicting solar irradiance at hourly and sub-hourly (10-minute) resolutions across Ireland. Cross-validation demonstrates that our model is statistically robust across all temporal resolutions with hourly showing highest prediction precision whereas 10-minute resolution encounters higher errors but better uncertainty quantification. In separate evaluations, we compare our model against alternative data sources, including reanalysis datasets and nearest-station interpolation, and find that it consistently provides superior site-specific accuracy. At the hourly scale, our model outperforms ERA5 in agreement with ground-based observations. At the sub-hourly scale,  10-minute resolution estimates provide solar PV power outputs consistent with residential and industrial solar PV installations in Ireland. Beyond surpassing existing datasets, our model delivers full uncertainty quantification, scalability and the capacity for real-time implementation, offering a powerful tool for solar energy prediction and the estimation of losses due to overload clipping from inverter undersizing.

\end{abstract}

\begin{highlights}

\item Novel spatio-temporal model using a Bayesian framework for estimating hourly and sub-hourly solar irradiance in Ireland with uncertainty. 

\item Comprehensive comparison of the spatio-temporal model for solar irradiance to reanalysis and ground-based data for site-specific solar PV modelling in Ireland.

\item Bias assessment reveals significant variability across data sources when estimating solar PV output in Ireland, with no single dataset consistently outperforming others under all conditions.

\item Preferred data source for predicting solar irradiance at any location in Ireland for hourly and sub-hourly time points is a Bayesian spatio-temporal model as it accounts for uncertainty and can provide near real-time estimates for solar irradiance.

\item Provides a method to quantify the extent of overload clipping at site level and the impact on commercial revenues.

\end{highlights}

\begin{keyword}
Spatio-temporal model \sep Bayesian framework \sep Uncertainty quantification \sep Solar irradiance \sep Ground-based measurements
\sep Reanalysis data \sep Solar photovoltaic data \sep overload clipping

\end{keyword}

\end{frontmatter}


\section{Introduction}
\label{introduction}

In today's rapidly changing climate, one of the primary challenges facing society is reducing our reliance on fossil fuels. A key strategy for decarbonising the energy grid is to increase the capacity for renewable energy sources, for example harnessing solar power. The Irish government has set bold targets to ensure 80\% of energy generation comes from renewable sources by 2030, including a target of 8 GW coming from solar power \citep{climate_action_plan2024}. In 2024, 40.1\% of all electricity generated in Ireland resulted from renewable electricity sources and of this total 2.1\% resulted from solar \citep{eirgrid_system_renewable25}. By February 2024, Ireland’s solar electricity generation capacity reached a milestone of 1GW
\citep{esb_solar1gw_2024}. It is clear a rapid increase in solar PV installations is required in order to reach ambitious targets. 

As the transition to a renewable energy based electricity system proceeds, it is important to understand the challenges associated with increasing the power generated by solar installations. At the electrical grid level, \cite{kerci2024emerging} outlined the four main challenges impeding an all-island power system for Ireland and Northern Ireland which included: dispatch down levels, long-term frequency deviations, voltage magnitude variations, and operational demand variations. Among these, dispatch-down, where available renewable energy cannot be used due to system-wide curtailments or localised constraints, is particularly significant \citep{eirgird2022_curtail}. In 2024, the total dispatch-down level from solar generation in Ireland was 5.3\% of total available solar production though this figure varied by month \citep{eirgrid_renewable_constraint_2024}. While several infrastructure improvements are planned to alleviate network constraints \citep[e.g.,][]{erigrid_soni2023}, other issues remain - such as limited quantification of the extent of residential rooftop solar contribution to the grid, and high frequency fluctuations and associated variability from large solar installations at high temporal resolution. One such additional phenomenon, known as overload clipping losses, arise when solar irradiance spikes drive inverters beyond their rated AC output \citep{villoz2022model}. Simulations that rely on hourly averaged data systematically underestimate these losses, since intra-hour irradiance peaks are smoothed out. Recent work has shown that hourly models can under-predict clipping losses by 1–5\% annually, especially in systems with high DC/AC ratios \citep{villoz2022model}. Sub-hourly modelling is therefore crucial, as it captures the short-term irradiance dynamics that directly affect inverter behaviour and system yield.

Addressing these challenges requires access to reliable solar irradiance data \citep[total solar power incident on a unit area of Earth:][]{solanki2013solar}, which is fundamental for estimating, assessing and forecasting energy generation from solar PV systems across Ireland. Solar irradiance or global horizontal irradiance (GHI) can be measured using observational data sources or it can be empirically estimated. There are a range of solar irradiance data sources which include ground-based measurements from national meteorological station networks \citep[e.g.][]{met_eireann_web}, reanalysis datasets such as ERA5 \citep{era5_paper} and MERRA-2 (Modern-Era Retrospective Analysis for Research and Applications) \citep{MERRA2paper}, and satellite data such as Surface Radiation Dataset Heliosat (SARAH) \citep{sarah_satellite_paper, SARAH_website}. Numerous countries have evaluated the most suitable solar irradiance data sources for quantifying the potential output of solar PV installations across different locations. For instance, \citet{YANG20203} provided a global-scale comparison between satellite- and reanalysis-derived GHI, while \citet{BRIGHT2019435} evaluated solar datasets from commercial providers such as Solcast \citep{solcast}. In the UK, \citet{palmer2018satellite} highlighted that the decision to use ground-based or satellite observations of solar irradiance depends on weather station density and the choice of satellite model, and can vary considerably across the country due to its highly changeable climate. Similarly, \citet{KENNY2022444} showed that significant seasonal and regional biases can arise in gridded irradiance datasets in Germany, reinforcing the need for local ground-based validation. Research in high-latitude regions like Norway has shown the added importance of accounting for factors such as snow cover to improve irradiance estimates \citep{NYGARDRIISE2024112975}. In South Africa, \citet{geomatics1040025} found that Solcast had the best agreement with ground-based data, while satellite data sourced from the Copernicus Atmosphere Monitoring Service \citep[CAMS:][]{copernicus_ads} and Satellite Application Facility on Climate Monitoring \citep[CMSAF][]{vzak2015cmsaf, pfeifroth2018satellite} were the most reliable freely available options. For Ireland specifically, \citet{su16156436} demonstrated that satellite-based irradiance data struggled to capture localised variability, leading to higher errors due to coarse spatial resolution.

While ground-based measurements are considered the most accurate due to their site-specific and real-time nature, their spatial coverage is sparse in Ireland, with just around 20 stations nationally. This limitation necessitates the use of gridded datasets to fill gaps in spatial coverage. Among reanalysis datasets, ERA5 has been recommended for Ireland due to its improved performance over MERRA-2 during summer months \citep{griffin2023climate, doddy2021reanalysis}. However, both reanalysis and satellite datasets suffer from key drawbacks: they are not available in real time, have limited sub-hourly temporal resolution and typically lack uncertainty quantification.

In this paper, we introduce a novel Bayesian spatio-temporal modelling framework for estimating solar irradiance across Ireland, specifically targeting locations without direct solar irradiance measurements. Unlike traditional methods such as nearest-neighbour interpolation \citep[e.g.][]{PEREZ199789} or kriging \citep[e.g.][]{YANG2018876}, our approach provides not only spatially continuous predictions but also the ability to make predictions in time at an hourly and sub-hourly resolution while fully quantifying uncertainty in predictions. This marks a significant advance in modelling solar resources for Ireland, and provides a framework that can be repeated in other countries. Using high-resolution spatial data, we assess how uncertainty in solar panel output varies across both space and time for sub-hourly intervals, an aspect that, to our knowledge, has not been previously explored in an Irish context. We also demonstrate how our models can serve as a scalable alternative for producing high-resolution solar resource estimates.

Section 2 introduces the data sources explored in this study. In Section 3, we describe the statistical modelling approach for solar irradiance in Ireland and discuss the mathematical formulation of solar irradiance on both flat and tilted surfaces. Section 4 presents the results of the model along with validation tests to assess performance. Finally, Section 5 offers conclusions and highlights key takeaways.

\section{Data}\label{datasets}
\begin{figure}[H]
    \centering
    \includegraphics[width=\textwidth]{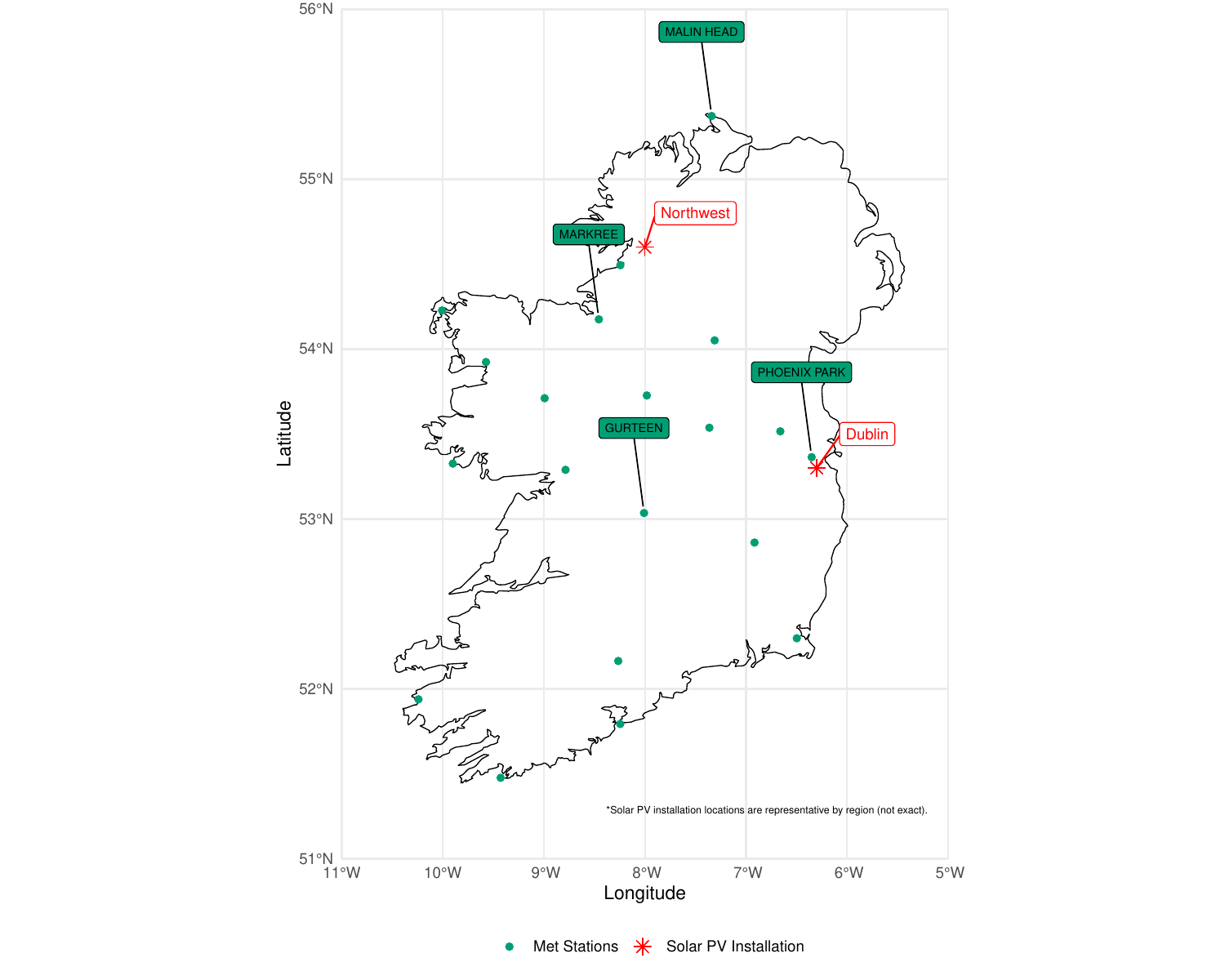}
    \caption{Locations of the study sites across Ireland. The ground-based meteorological stations are shown as green circles and are sourced from \citet{met_eireann_web}. The four case study sites selected for detailed analysis are labelled in green. The two solar PV installation sites are represented by red stars, with locations depicted at a regional level (Northwest and Dublin) and not the exact coordinates, in order to preserve data privacy.}
    \label{fig:map_data_sources}
\end{figure}

In this paper, we examine a range of data sources which include ground-based measurements from meteorological (met) stations, reanalysis datasets and two solar PV installation datasets. Figure \ref{fig:map_data_sources} presents the locations of each data source used in this paper which are discussed in the upcoming section. 

\subsection{Pyranometer in meteorological stations}\label{met_data_info}
In the Republic of Ireland, ground-based solar irradiance measurements are obtained using pyranometers (sensors that measure solar radiation), which record data at a one-minute temporal resolution \citep{met_eireann_web}. These instruments are deployed at 20 meteorological stations across the country, as illustrated in Figure \ref{fig:map_data_sources}. Each station is managed and maintained by Met Éireann, the national meteorological service of Ireland. The data collected from these stations are widely regarded as the most reliable and high-quality source of solar irradiance information available in Ireland. However, the spatial coverage of the network is limited, particularly in the south-western region of Ireland. Recent findings by \citet{griffin2023climate} have highlighted the absence of pyranometer measurements at key locations such as Shannon and Cork airports, as well as issues with missing data records at existing sites. Given the sparsity of observations in some areas, the report advises against interpolating across large spatial regions, as this may yield unreliable estimates. Additionally, data from Northern Ireland was excluded from this analysis due to the lack of publicly available, real-time data at sub-hourly resolution, however we do provide estimates for solar irradiance for the whole island.

For illustrative purposes, we have selected four meteorological stations for a single date in January and June 2024 (Figure \ref{fig:raw_data_4_site}) which we use to demonstrate our findings and the remaining sites are plotted in the Appendix. These plots indicate large variability in solar irradiance across the day and at the different locations. The raw data collected at the meteorological stations is at a 1 minute resolution, however we aggregate this to 10-minute and hourly resolution to align with other data sources used for validation purposes. By aggregating 1 minute data over the hour, we see dramatic smoothing at each location demonstrating the lose of 98\% of the raw data points for 2024. As a result, we lose the ability to detect minute-to-minute fluctuations and short solar irradiance peaks are smoothed out or missed. However, modelling at a 1 minute resolution results in extensive computational requirements, therefore 10-minute and hourly resolution is utilised (Figure \ref{fig:raw_data_4_site}). The 10-minute resolution data has less variability compared with the 1 minute resolution data, yet possesses significantly more fluctuations compared with the hourly resolution highlighting the short term changes in solar irradiance. 

\begin{figure}[H]
    \centering
    \includegraphics[width=\linewidth]{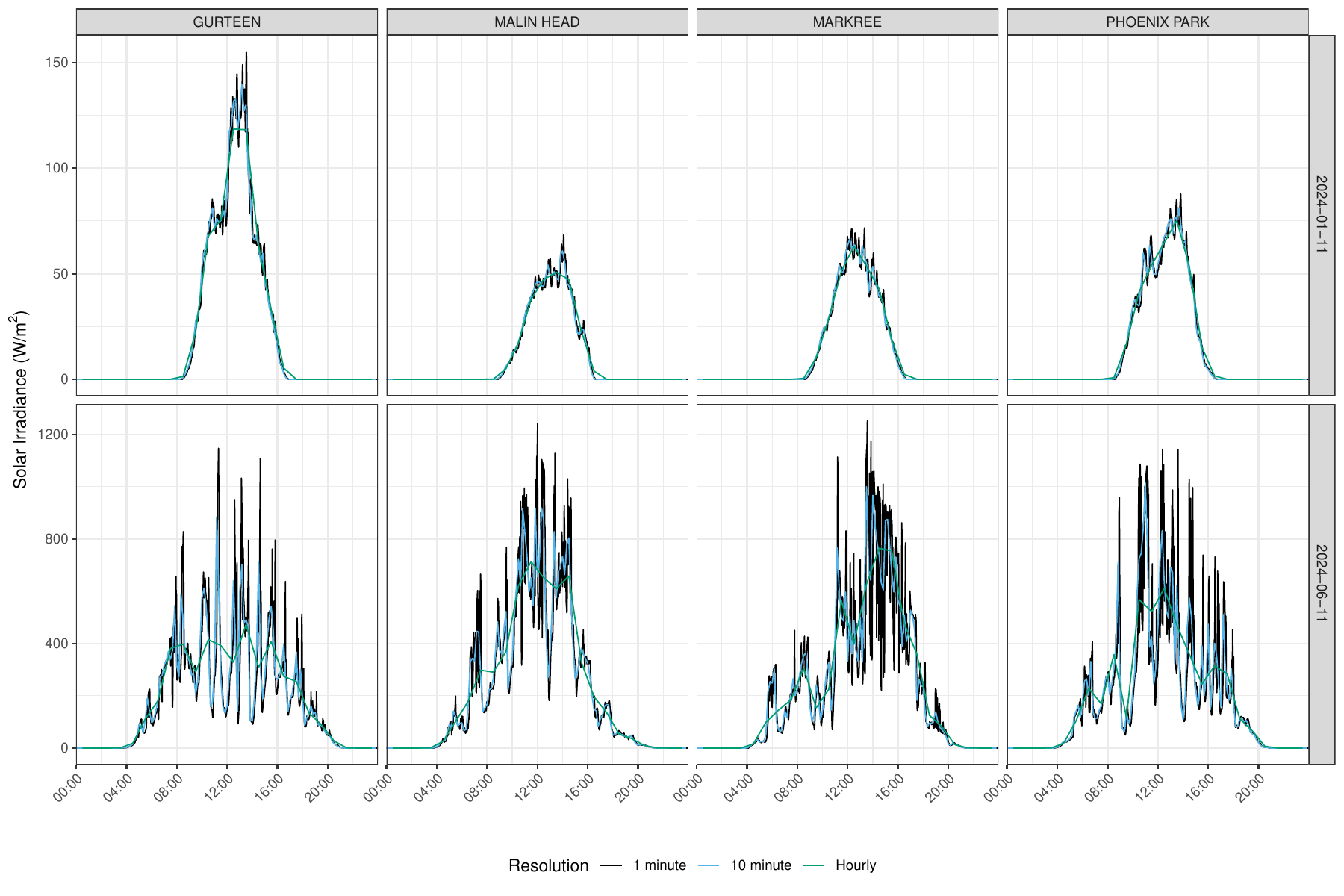}
    \caption{Plot of solar irradiance at 1 minute, 10-minute and hourly resolution from 4 meteorological stations weather stations for 11th January 2024 and 11th June 2024 \citep[data provided by ][]{met_eireann_web}.}
    \label{fig:raw_data_4_site}
\end{figure}

\subsection{Reanalysis Data}
Reanalysis data is formed by retrospectively analysing historic climate conditions using advanced data assimilation techniques \citep[e.g.,][]{beng2004}. In this paper, we focus on one popular reanalysis product; ERA-Interim reanalysis of the European Center for Medium-range Weather Forecasts \citep[ECMWF:][]{era5_paper}. This reanalysis dataset has high spatial resolution with ERA5 grid resolution of $0.25^{\circ}\times 0.25^{\circ}$. However, it is limited to hourly temporal resolution, is not available in real time, and provides no uncertainty quantification for the solar irradiance estimates. We note that other reanalysis datasets are popular, e.g. the MERRA-2 (Modern-Era Retrospective Analysis for Research and Applications) reanalysis dataset proposed by the NASA Global Modelling and Assimilation Office \citep{gelaro2017modern} and the Irish Met Éireann reanalysis simulation  \citep[MÉRA;][]{gleeson2017met}. However, \citet{mathews2023systematic} demonstrated that ERA5 possesses systematic bias when simulating solar PV outputs, albeit these biases appear smaller than those of MERRA-2 for the cloudy climate of Northern Ireland. For the MÉRA dataset, the model tends to have a negative bias when forecasting irradiance on cloudy days and the reanalysis dataset is limited to a 35 year period ending in 2019 \citep{atmos9050163}. Therefore, we focus on the ERA5 dataset for validations of our models for the hourly resolution case and we do not incorporate them into the model as an additional data source. For ERA5 a simple nearest point interpolation for 2024 is used for model validations.

\subsection{Solar photovoltaic installation data}\label{solar_pv_data}
As an additional validation dataset, we have one residential solar PV system and one industrial solar PV system. The residential system is installed on the roof of a house in Dublin and comprises of seven panels rated at 400 W each, paired with a 2,450 W inverter. For this system, we have hourly data covering the full year of 2024, along with higher-resolution data at 10-minute intervals for two representative months, January and June. 

The industrial system is located in the north-west region of Ireland and has 100 panels with DC-to-AC ratio of approximately 1.2 (ratio of inverter size to potential max output: \citep{KAEWNUKULTORN2024121402}). For the industrial system, data are available for January and June, at 10-minute resolution. For both systems, we have manually recorded key structural characteristics, including panel tilt and orientation. Compared to pyranometer measurements, the PV systems exhibit additional sources of noise due to its physical positioning and varying levels of system maintenance, with no available information on cleaning frequency or fault monitoring. In accordance with data consent agreements and to protect the confidentiality of system owners, detailed information about the PV systems will not be disclosed. The data have been aggregated to the regional level to preserve location anonymity, with the red stars in Figure \ref{fig:map_data_sources} representing illustrative locations.

\section{Methodology}
In this section, we present our novel statistical model to provide estimates for solar irradiance for any location in Ireland at any historical time point. We also provide the mathematical formulation of solar irradiance for both flat and tilted surfaces which are used to estimate potential solar PV installation output.

\subsection{Bayesian Spatio-Temporal model for solar irradiance in Ireland}\label{model_desc}
Bayesian statistics is a method of data analysis that updates prior beliefs about model parameters using observed data through Bayes’ theorem to produce a posterior distribution for inference and prediction, offering key advantages such as incorporating prior knowledge, fully quantifying uncertainty and enabling flexible modelling in complex or data-scarce situations \citep{intro_bayes2021}. Computational methods are used to approximate the posterior distribution when it cannot be calculated exactly; common techniques include traditional Markov Chain Monte Carlo (MCMC) methods \citep{gilks1995markov} and the Integrated Nested Laplace Approximation \citep[INLA,][]{rue2009approximate}. INLA offers a faster alternative for approximate Bayesian inference by leveraging Gaussian Markov random fields (GMRFs) \citep{held2009posterior} which is why it is chosen for this paper \citep[for more details on these methods refer to;][]{rue2017bayesian, bakka2018spatial, krainski2018advanced}. We use the R-library INLA (\url{www.r-inla.org}) in our analysis.

When developing a Bayesian hierarchical spatio-temporal model for solar irradiance in Ireland, we use solar irradiance data from 20 metrological stations across Ireland supplied by Met Éireann \citep{met_eireann_web} which is recorded at a 1-minute temporal resolution. In this paper, we highlight two potential modelling scenarios; hourly solar irradiance and 10-minute solar irradiance. For the modelling scenarios, we aggregate the 1-minute data to hourly solar irradiance and sub-hourly solar irradiance separately. Additionally, due to the presence of zeros in our dataset, corresponding to night-time, we transform solar irradiance to be modelled using the log of solar irradiance plus 1. 

We define
$ y_{i}(\mathbf{x}, t) = \log\!\big(\text{irradiance}(\mathbf{x}, t) + 1\big)$, where irradiance is measured in watts per square meter ($W\,m^{-2}$), at location $\mathbf{x}$, time $t$ and day of the year $i$. Each day of the year is modelled independently to ensure feasible computational run-times and to account for the discontinuities at night, which would otherwise require zero-inflated formulations \citep[e.g.][]{feng2021comparison}. These daily models also align with the horizon of energy market operations in Ireland \citep{SEMOpx2025}, which include day-ahead and intra-day trading structures, while allowing exploration of finer temporal resolutions within each day for potential real-time estimates. Additionally, our external validation datasets (described in Section \ref{datasets}) ranged in length from 1 month to 1 year, allowing the model to efficiently examine statistical properties for specific days. The model is specified as:
\begin{equation}\label{eq:model_eq}
    y_{i}(\mathbf{x},t) = \beta_0 + u(\mathbf{x},t) + \epsilon_{i}
\end{equation}
where $\beta_0$ is the overall intercept or the average level of solar irradiance response across space and time. The error vector is assumed to be normally distributed with independent identically distributed (i.i.d) $\epsilon_{i} \sim N(0,\sigma_e^2)$, and captures any additional variation not accounted for in the spatial component or the error associated with the observation. $u(\mathbf{x},t)$ is a latent space-time Gaussian field representing the underlying spatio-temporal variation in the response. The Gaussian assumption makes inference tractable since the field is determined only by a mean, variance, and covariance function, rather than having to estimate a value at every location and time independently \citep{williams2006gaussian}. To balance computational efficiency with model complexity, $u(\mathbf{x},t)$ is modelled using a separable space-time structure, where the joint space-time correlation structure is decomposed into a temporal and a spatial component \citep{cameletti2011comparing, krainski2018advanced}.

The temporal dependence of the latent spatio-temporal Gaussian field $u$ is modelled using an autoregressive process of order 1 (AR(1)). Specifically, the temporal correlation structure of $u$ at a fixed spatial location $\mathbf{x}$ is given by:
\[
C(u(\mathbf{x},t),\, u(\mathbf{x},t')) = \rho^{|t - t'|}, \qquad |\rho| < 1.
\]

where $\rho$ is the lag-1 correlation parameter describing the strength of correlation between consecutive time points. This formulation ensures that the temporal correlation decays geometrically as the temporal separation $|t - t'|$ increases, meaning that time points closer together are more strongly correlated than those further apart.

For the spatial structure of the latent spatio-temporal Gaussian field $u$, the dependence between two locations $\mathbf{x}$ and $\mathbf{x}'$ is described by the Matérn correlation function \citep{williams2006gaussian}:
\[
C(u(\mathbf{x},t),u(\mathbf{x}',t))
    = \sigma^2 \frac{1}{2^{\nu-1}\Gamma(\nu)}\,
      \big(\kappa \|\mathbf{x}-\mathbf{x}'\|\big)^\nu 
      K_\nu\!\big(\kappa \|\mathbf{x}-\mathbf{x}'\|\big).
\]
where $\|\mathbf{x}-\mathbf{x}'\|$ is the distance between sites and $\sigma^2>0$ is the variance associated with the spatial process. $\nu>0$ controls the smoothness of the field and is fixed to 1 within the INLA framework for improved computational efficiency \citep{gomez2020bayesian}. $\kappa>0$ controls how quickly correlations decay with distance (small $\kappa$ means locations remain correlated over larger distances), meaning that $u(\mathbf{x},t)$ varies smoothly in space and evolves over time with a memory that decays as observations are further apart. It is common practice to work with the effective spatial range $\phi$ which is given by $\phi = \sqrt{8}/\kappa$ \citep{gomez2020bayesian}. $K_{\nu}$ is the modified Bessel function of second order \citep{williams2006gaussian}. 

\begin{figure}[H]
    \centering
    \includegraphics[width=0.8\linewidth]{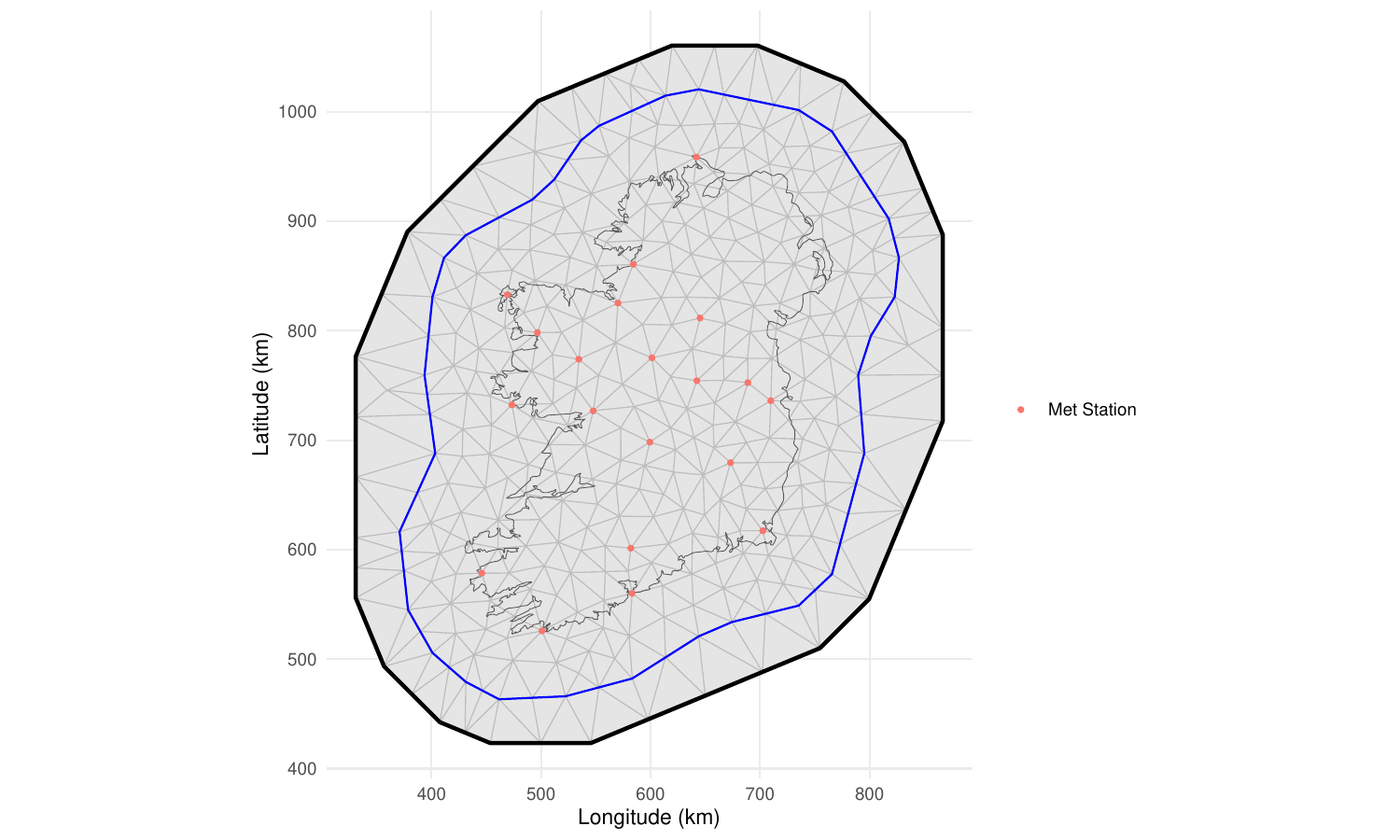}
    \caption{The mesh to solve the stochastic partial differential equation for the spatial field representing Ireland. We include red points to represent the meteorological stations (Met Stations). The mesh has two distinct zones an inner zone which has a higher resolution as it contains the required spatial locations of our Met station. The second zone in this case is coarser in order to reduce boundary effects of the mesh. There are 362 nodes in this mesh. We include a map of Ireland to visualise how the mesh corresponds to Ireland. }
    \label{fig:mesh}
\end{figure}

We represent the spatial component of our Gaussian field using a Gaussian Markov Random Field (GMRF), which represents a spatial field as a network of connected points or nodes, where each node is influenced only by its neighbours \citep{krainski2018advanced}. To approximate the continuous spatial field, we implement the GMRF via a stochastic partial differential equation (SPDE) approach \citep{lindgren2011explicit}. The SPDE approach converts the continuous surface into a triangulated mesh of nodes (see Figure  \ref{fig:mesh}), allowing us to capture spatial variation accurately while keeping computations feasible for large datasets.

The mesh (Figure \ref{fig:mesh}) has 362 nodes, with a finer resolution around the meteorological stations (red dots) to accurately represent the spatial locations of our data, and a coarser resolution near the boundaries to reduce computational cost and avoid boundary effects. The mesh is used to construct a sparse precision matrix, which further enhances computational efficiency. A sparse precision matrix allows for faster computation because the matrix contains mainly zeros except for when the nodes of the mesh (shown in Figure \ref{fig:mesh}) have non-zero connections.

Within our Bayesian framework, we provide priors for our parameters of interest. The parameters of interest include the intercept $\beta_0$, the  spatio-temporal latent field $u(\mathbf{x},t)$ and the observation error variance $\sigma^2_{\epsilon}$. 
The spatio-temporal latent field captures deviations in solar irradiance across space, with the spatial standard deviation $\sigma$ reflecting variability between locations, the effective range $\phi$ indicating the distance over which solar irradiance is correlation, and the temporal AR(1) correlation $\rho$ measuring persistence over time. The prior for spatial parameters $\phi$ and $\sigma^2$ are given penalized complexity priors \citep{fuglstad2019constructing}. The remaining components of the model, including $\beta_0$, $\rho$ and $\sigma_{\epsilon}$, use the default prior settings provided by \texttt{INLA}. 

\subsection{Empirical model for solar irradiance}
The three main components of solar irradiance for a flat surface are; the Global Horizontal Irradiance (GHI), Beam Horizontal Irradiance (BHI) and Diffuse Horizontal Irradiance (DHI) \citep{Perpinan2023}. Commonly, GHI is measured by ground-based pyranometers, while the remaining components are less regularly recorded due to cost and complexity, and instead require an empirical model for estimates \citep{NUNEZMUNOZ2022122820}. In this paper, we present a spatio-temporal model using pyranometers in Ireland to represent GHI and using the \texttt{solaR} package \citep{solaR_package} to estimate the remaining components. Other packages exist, for example \texttt{pvlib}\citep{anderson2023pvlib} - we utilised \texttt{solaR} due to its compatibility with our other models which are coded in the R programming language. 

The components of solar irradiance are given as follows:
\begin{equation}
    GHI = BHI + DHI
\end{equation}
where $GHI$ is the total solar irradiance received on a horizontal surface, including both direct and diffuse components. The $BHI$ is the direct normal irradiance (DNI) projected onto a horizontal surface given by:
\begin{equation}
    BHI = DNI * cos \theta_z
\end{equation}
where $DNI$ is the direct normal irradiance (irradiance received per unit area normal to the sun's rays) and $\theta_z$ is the solar zenith angle (angle between the sun and the vertical direction). As DHI and DHI are not recorded, a decomposition model is used to calculate them. In order to estimate the $DNI$ component, \texttt{solaR} uses the \citet{ridley2010modelling} technique. 

$DHI$ is the scattered radiation received from the sky and can be expressed by substituting for BHI:
\begin{equation}
    DHI = GHI - BHI = GHI - (DNI * cos\theta_z)
\end{equation}
In other words, the DHI is the residual part of GHI after removing the direct component also known as a decomposition model \citep{Perpinan2023}. Alternatively, it is possible to estimate DHI using empirical models however, this is beyond the scope of this paper.

The solar zenith angle, $\theta_z$, is the angle between the sun and the vertical direction and is calculated by:
\begin{equation}
    \theta_z = 90^{\circ} - e
\end{equation}
where $e$ is the elevation angle given by:
\begin{equation}
    \sin(e) = \sin(\phi)\sin(\delta) + \cos(\phi)\cos(\delta)\cos(h)
\end{equation}
where $\phi$ is the latitude, $\delta$ is the solar declination angle and $h$ is the hour angle. The solar declination angle depends on the day of the year and is related to the sun's orbital motion. The hour angle represents how far the sun has moved from solar noon. In order to calculate these solar geometric parameters, we use the default setting in the \texttt{solaR} package which implements the \citet{michalsky1988} approach which we summarise in this section.

In summary, the GHI includes contributions from both the direct solar radiation projected onto the horizontal plane and the diffuse radiation scattered from the atmosphere summarised as:
\begin{equation}
    GHI = DNI* cos\theta_z + DHI
\end{equation}

An alternative relationship is required when examining solar PV installations. The plane-of-array irradiance $G_{ef}$ is obtained from the solar irradiance incident on the tilted plane of the PV modules \citep{Perpinan2023}. It is derived from the decomposition and transposition \citep[e.g.,][]{DRIESSE2024112093} of global horizontal irradiance (GHI), considering the direct, diffuse and reflected components described by:
\begin{equation}
    G_{ef} = G_b + G_d + G_r
\end{equation}
where $G_b$ is the beam (direct) irradiance on the tilted plane, $G_d$ is the diffuse irradiance on the tilted plane and $G_r$ is the ground-reflected irradiance. 

The direct component on a tilted surface, such as a solar PV panel, can be calculated using the direct solar radiation on a horizontal plane (the pyranometer) using trigonometry \citep{RIDLEY2010478}. The direct solar beam is given by \citep{maxwell1986measuring}: 
\begin{equation}
    G_b = DNI * cos \theta
\end{equation}
where $DNI$ is the direct normal irradiance and $\theta$ is the angle of incidence (angle between the sun's rays and the module normal). Angle of incidence $\theta$ is the angle between normal of a surface and a beam of radiation incident on it. For horizontal surfaces, like the pyranometer used by Met station, the incident angle and the the zenith angle are the same \citep{Perpinan2023}. To calculate the angle incidence we use the following:
\begin{equation}
\begin{split}
    cos \theta = sin(\delta)*sin(\phi)*cos(\beta) -\\ sin(\delta)*cos(\phi)*sin(\beta)*cos(\gamma)+\\
    cos(\delta)*cos(\phi)*cos(\beta)*cos(h)+ \\ cos(\delta)*sin(\phi)*sin(\beta)*cos(\gamma)*cos(h) +\\
    cos(\delta)*sin(\beta)*sin(\gamma)*sin(h)
\end{split}
\end{equation}
where $\delta$ is the solar declination angle, $\phi$ is the latitude, $\beta$ is the tilt of the panel, $\gamma$ is the azimuth angle of the panel and $h$ is the hour angle. 
 
The diffuse irradiance component is estimated using an anisotropic sky model such as \citet{PEREZ199789} given by:
\begin{equation}
    G_d = DHI * F
\end{equation}
where DHI is the diffuse horizontal irradiance (diffuse part of GHI) and $F$ is a function that models the diffuse transposition. A common approximation using the isotropic model \citep[e.g.,][]{duffie2013} is:
\begin{equation}
    G_d = DHI *\frac{1+cos\beta}{2}
\end{equation}
where $\frac{1+cos\beta}{2}$ represents the fraction of the sky dome visible to the tilted surface. For anisotropic models, for example \citet{davies1980calculation} approach, additional terms account for circumsolar and horizon brightness which are the default model used by \texttt{solaR}. It is formulated in the following manner:
\begin{equation}
    G_d = DHI (F_1 \frac{cos(\theta_i)}{cos(\theta_z} + (1-F_1)\frac{1+cos(\beta)}{2})
\end{equation}
where $\theta_i$ is the incident angle, $\theta_z$ is the zenith angle and $F_1$ is the anisotropy index (based on DNI/DHI ratio).

The ground-reflected component is given by:
\begin{equation}
    G_r = GHI * \rho * \frac{1- cos \beta}{2}
\end{equation}
where $\rho$ is the ground albedo (typically 0.2 for grass, higher for snow) and $\frac{1-cos\beta}{2}$.

In summary, the equation to transform a horizontal to tilted plane irradiance is given by:
\begin{equation}
    G_{ef} =  DNI * cos\theta + DHI * F + GHI * \rho * \frac{1-cos\beta}{2}
\end{equation}

\subsection{Implementation of software}
The spatio-temporal surface for solar irradiance is formed using the model structure described in equation \ref{eq:model_eq} using \texttt{INLA} and a separate model is developed for each time resolution; 10-minutes and hourly. The various time resolutions were chosen to correspond with the temporal resolution of our validation datasets and computational capacity. To run the hourly spatio-temporal model for 366 days takes 4 hours and 48 minutes. Therefore, it takes approximately 1 minute to run each day individually. For 10-minute resolution data the model takes approximately 10 minutes to run each day individually.

Next, we take our spatial temporal surface, at hourly and sub-hourly temporal resolutions for Ireland, and use them as inputs in the \texttt{solaR} package \citep{solaR_package} in order to calculate solar PV generation. Within the \texttt{solaR} package, the function to calculate the energy production of a Grid-Connected Photovoltaic (GCPV) system is implemented \citep{perpinan2012solar}. This function uses the photovoltaic energy balance, considering the irradiance, efficiency losses and system parameters.

\section{Results}
In this section, we present our spatial temporal maps for hourly and sub-hourly solar irradiance for Ireland. We highlight model performance using leave-one-site-out cross validation technique. Following our model performance analysis, we compare our estimated solar irradiance with ground-based measurements and ERA5 reanalysis dataset. Additionally, we demonstrate our model performance for solar power estimates using the \texttt{solaR} package \citep{solaR_package} and compare our results to a solar PV system at hourly and sub-hourly resolution. In Table \ref{tab:data_models_list} a summary of the model type, the input data, the model validation set and the external validation set is listed. 
\begin{table}[H]
    \centering
    \begin{tabular}{|| p{4cm} | p{4cm} | p{3cm} | p{4cm}| p{4cm}||} 
 \hline
Model Type& Input Data & Model Validation Technique & Validation Data Source  \\ [0.5ex]
\hline
         Hourly spatio-temporal model & Hourly Met Station Data & Leave-one-site-out & ERA5   \\ 
         \hline
         10-minute spatio-temporal model & 10-minute Met Station Data & Leave-one-site-out & 2 Solar PV installations \\
         [0.5ex]\hline
    \end{tabular}
    \caption{Overview of the spatio-temporal models for solar irradiance developed in this study, including their temporal resolution, input datasets for training, the internal validation technique applied and the external datasets used for model evaluation. The hourly model is trained on hourly meteorological station data and internally validated using leave-one-site-out cross-validation, with model performance externally compared against ERA5 reanalysis data. The 10-minute model is trained on 10-minute meteorological station data, internally validated with leave-one-site-out, and externally evaluated against measurements from two solar PV installations.}

    \label{tab:data_models_list}
\end{table}

\subsection{Hourly spatio-temporal surface of solar irradiance in Ireland}
\begin{figure}[H]
    \centering
    \includegraphics[width=\linewidth]{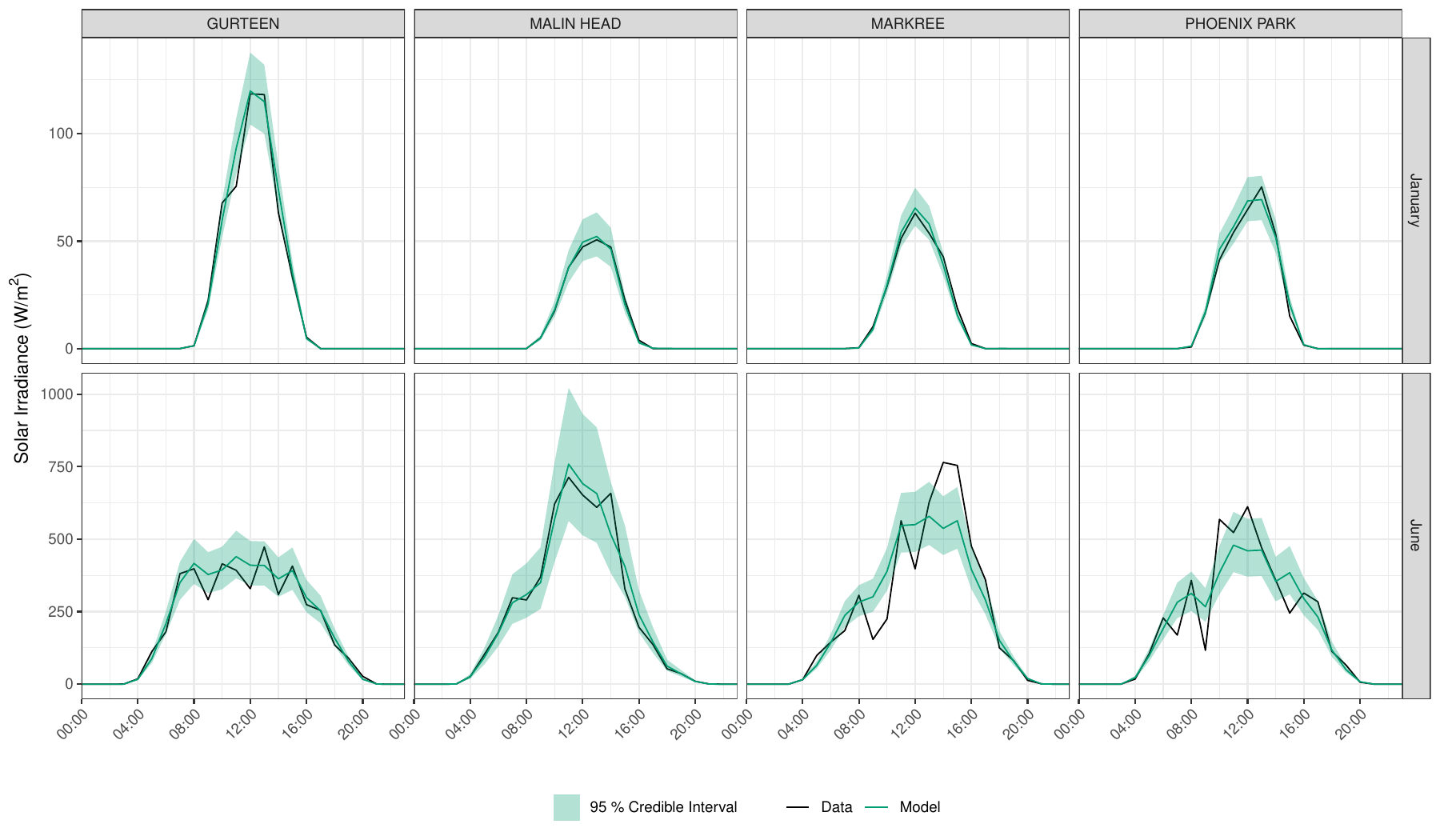}
    \caption{Time series plot for 11th January and 11th June 2024 for hourly resolution using our four case study locations. The raw data provided by Met Éireann is given in black and the model fit along with the 95\% credible interval is given in green. The x-axis is the solar irradiance recorded at an hourly temporal resolution. }
    \label{fig:timeseries_hourly_4sites}
\end{figure}

In this section, we present results from the Bayesian spatio-temporal model for hourly solar irradiance in Ireland. The model uses input data from 20 Met Éireann stations, aggregated to hourly resolution. Figure \ref{fig:timeseries_hourly_4sites} shows two representative days, 11 January and 11 June 2024, for our four case study sites, with 95\% credible intervals indicated by shading. Irradiance is much higher in June (0–1000 W/m$^2$) than in January (0–140 W/m$^2$), and January profiles are smoother. The model captures January patterns well and performs strongly in June for Gurteen and Malin Head, though additional peaks at Markree are not fully represented, highlighting spatial variation in irradiance across Ireland. As noted in Section \ref{met_data_info}, hourly aggregation smooths short-term fluctuations, motivating our extension to sub-hourly modelling.

\begin{figure}[H]
    \centering
    \begin{subfigure}[b]{0.48\linewidth}
        \centering
        \includegraphics[width=\linewidth]{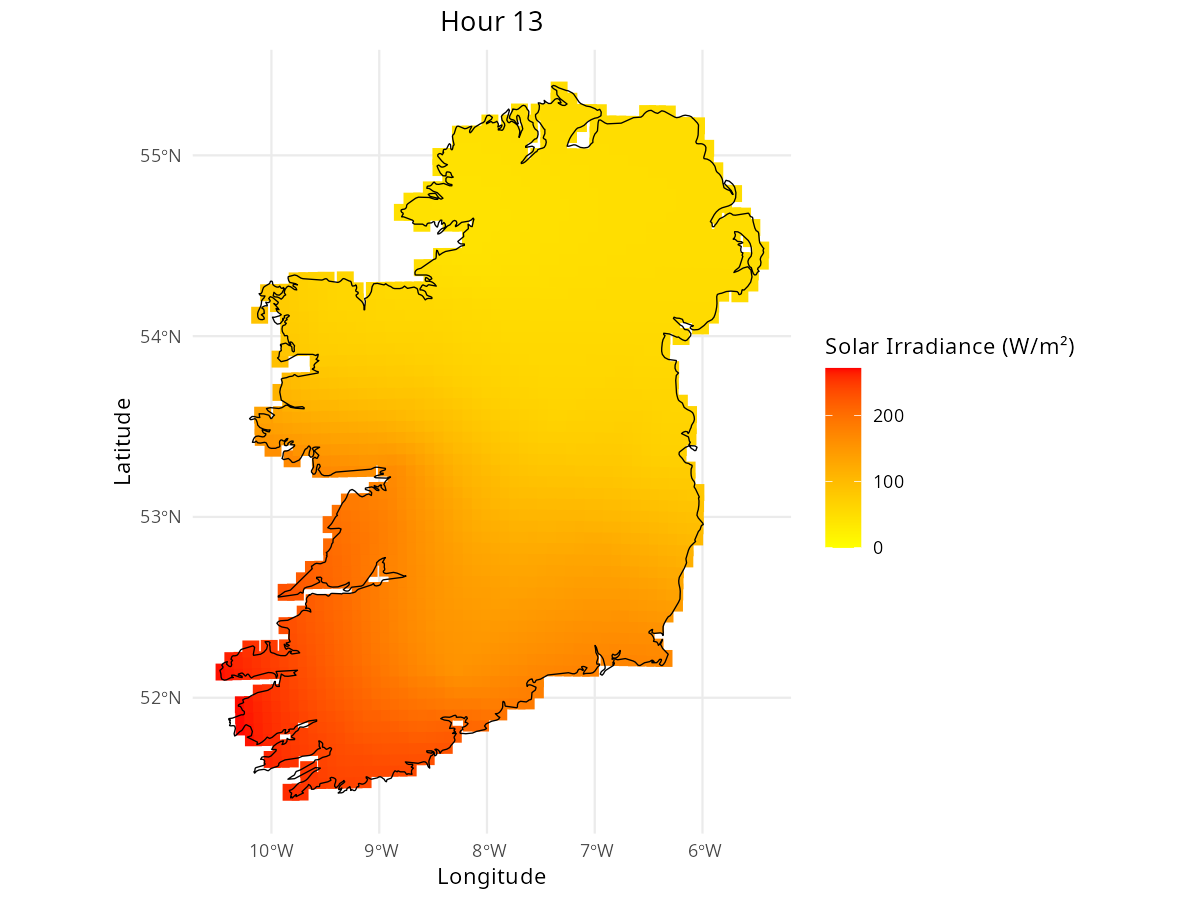}
        \caption{Posterior mean predicted solar irradiance for Ireland in W/m$^2$ on 11th January 2024 at 1pm}
        \label{fig:surf_hour_jan_ni}
    \end{subfigure}
    \hfill
    \begin{subfigure}[b]{0.48\linewidth}
        \centering
        \includegraphics[width=\linewidth]{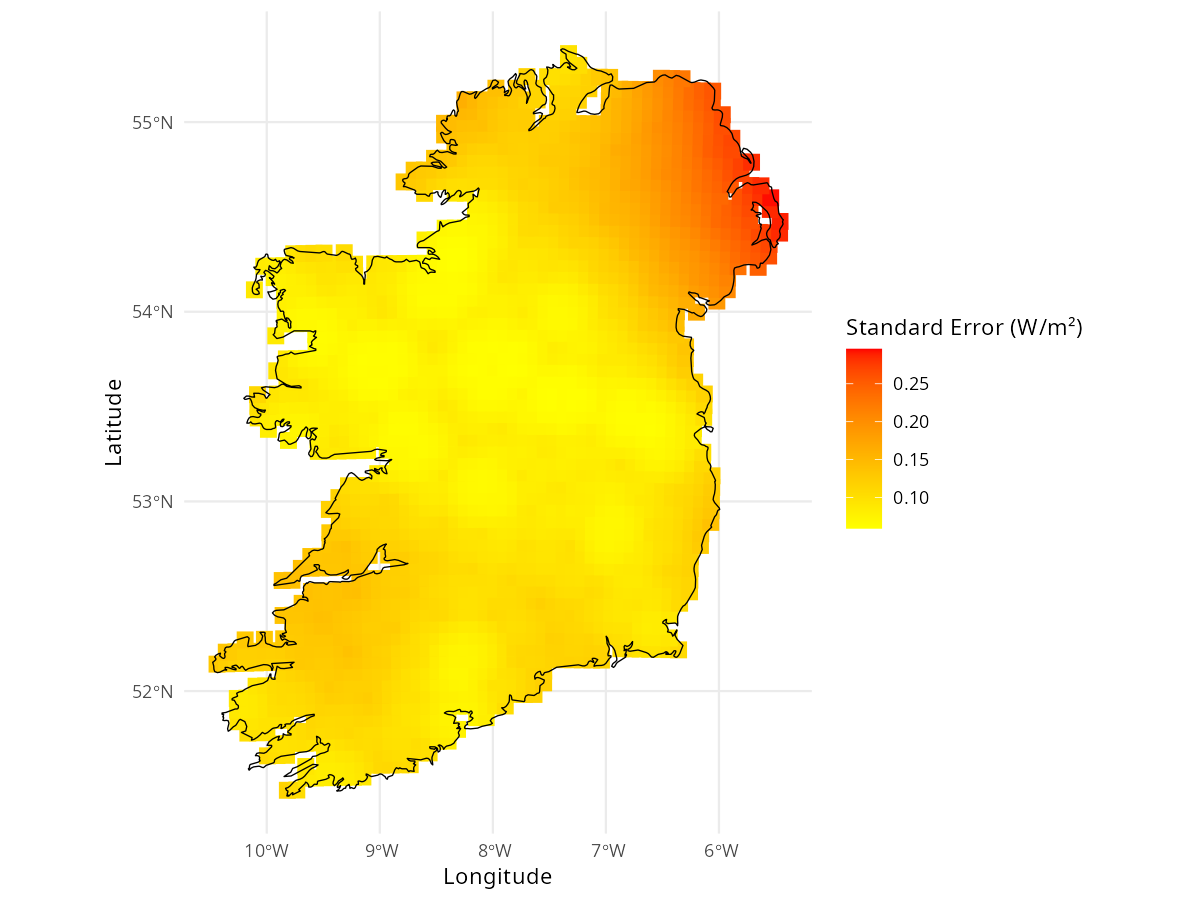}
        \caption{Standard error of posterior mean predicted solar irradiance for Ireland in  W/m$^2$ on 11th January 2024 at 13:00}
        \label{fig:surf_hour_jan_error}
    \end{subfigure}
    \begin{subfigure}[b]{0.48\linewidth}
        \centering
        \includegraphics[width=\linewidth]{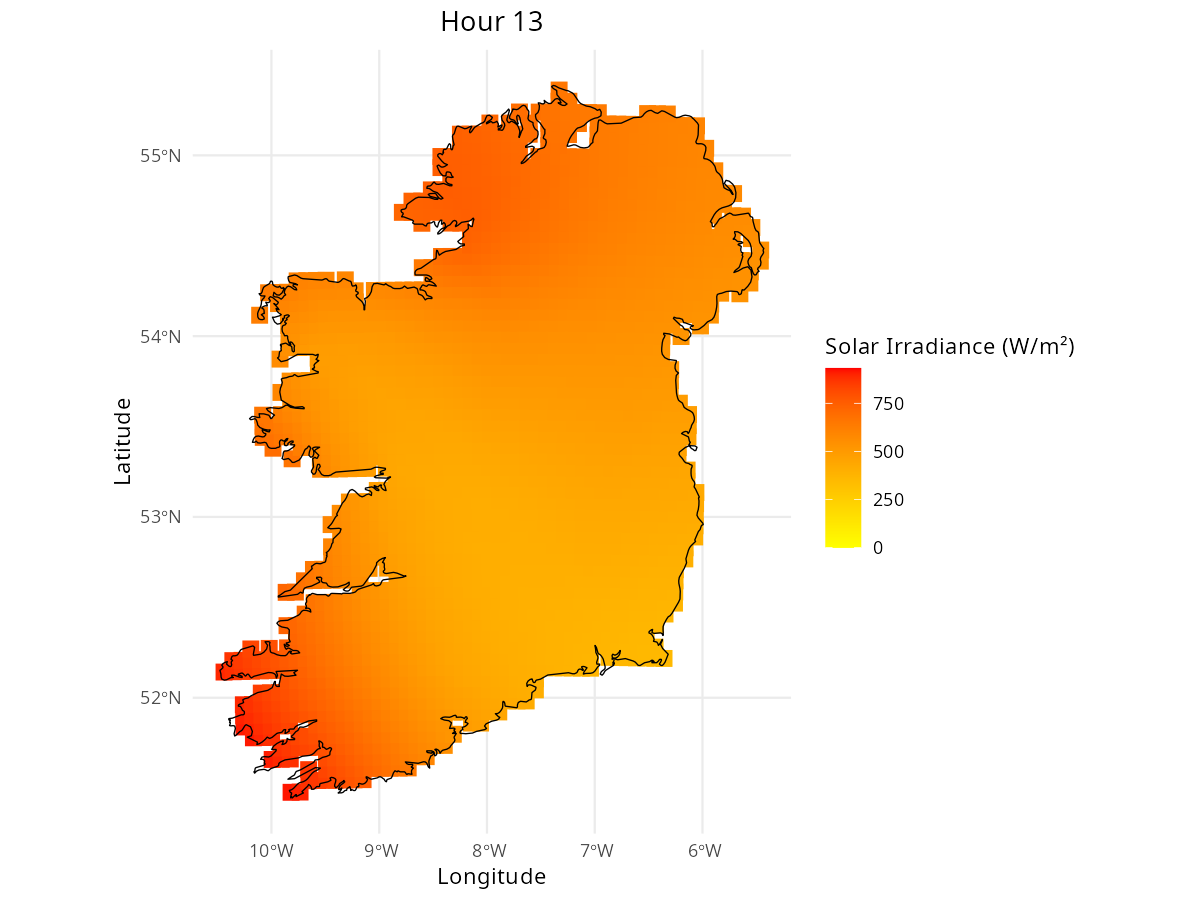}
        \caption{Posterior mean predicted solar irradiance for Ireland in W/m$^2$ on 11th June 2024 at 1pm}
        \label{fig:surf_hour_june_10}
    \end{subfigure}
    \hfill
    \begin{subfigure}[b]{0.48\linewidth}
        \centering
        \includegraphics[width=\linewidth]{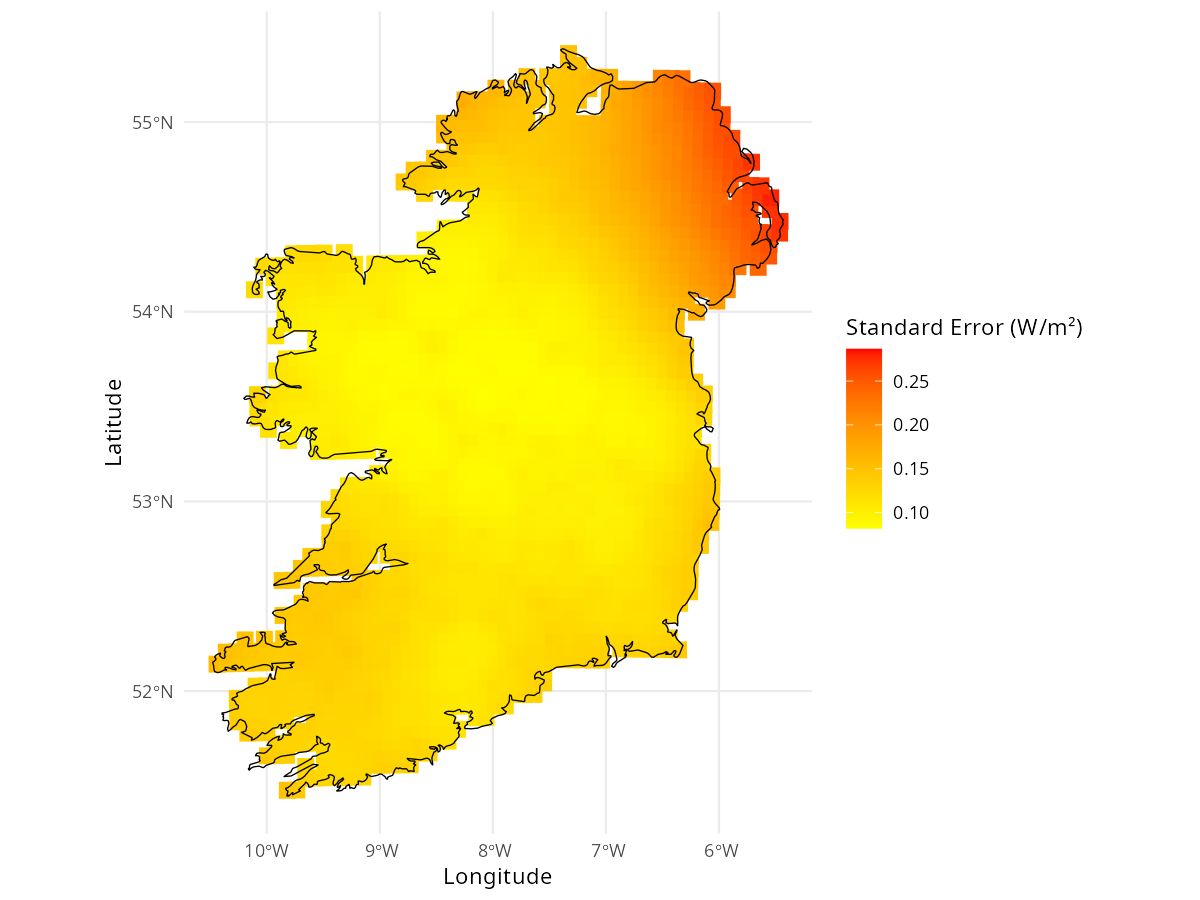}
        \caption{Standard error of posterior mean predicted solar irradiance for Ireland in  W/m$^2$ on 11th June 2024 at 13:00}
        \label{fig:surf_hour_june_10_error}
    \end{subfigure}

    \vspace{0.5cm}
    \caption{Posterior mean solar irradiance and associated standard errors across Ireland at 13:00 (1 pm) as estimated by a spatio-temporal Bayesian INLA model for January and June 2024. Each row corresponds to a month, with the left panels (a,c) showing the predicted solar irradiance surface (W/m²) and the right panel (b,d) showing the corresponding standard errors. The black outlines indicate the border of Ireland and Northern Ireland, highlighting spatial variation and model uncertainty in hourly solar irradiance.}
    \label{fig:surf_hour_all}
\end{figure}

In Figure \ref{fig:surf_hour_all} we present maps of Ireland for 1pm on the 11th of January (a) and June 2024 (c). In January the range of solar irradiance is significantly smaller, 0 and 220 W/m$^2$, compared to the solar irradiance range in June of 0 and 800 W/m$^2$. The associated standard error estimates for each location from the model is presented for January (b) and June (c) and possesses similar ranges in values from 0.1 to 0.25 W/m$^2$. The standard error at each location provides an insight into how precise the model prediction is and relative to the size of the solar irradiance, the standard errors are small. Yet, in certain locations, there is substantial uncertainty, particularly for Northern Ireland where there is no input data included in the models as the data is not available. 

\subsection{10-minute spatio-temporal surface of solar irradiance in Ireland}
To further demonstrate our flexible model structure, we present results for our 10-minute spatio-temporal approach. As previously mentioned, we aggregate the 1 minute meteorological station data for 20 stations to 10-minute intervals and use this as input data for our models. At this temporal scale, we can reduce the amount of solar irradiance information lost by smoothing the data to hourly levels. We can also compare our results with a separate validation dataset from two solar PV installations, which are recorded at 10-minute temporal interval, facilitating direct comparison and validation.

\begin{figure}[H]
    \centering
    \includegraphics[width=\linewidth]{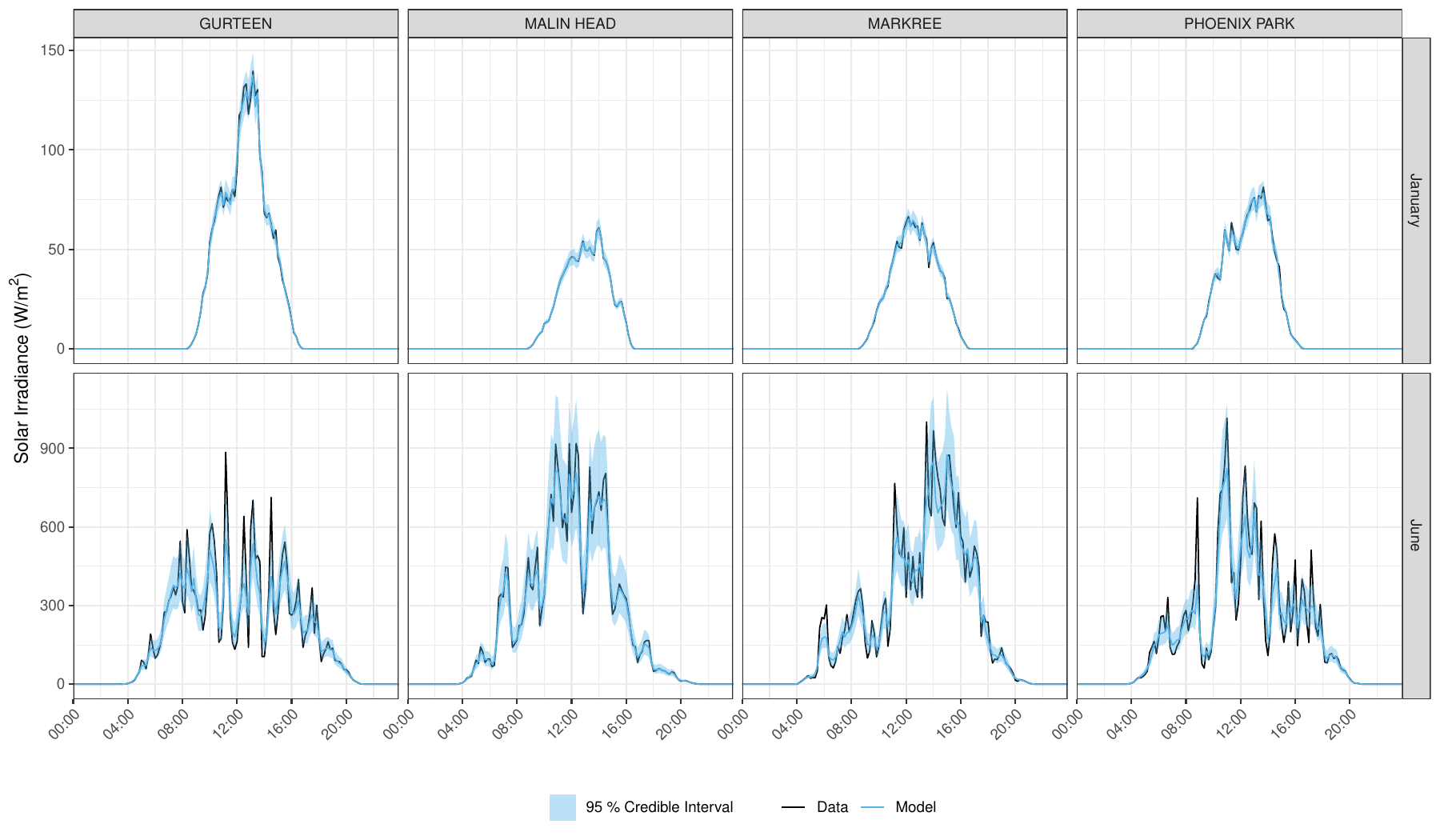}
    \caption{Time series plot for 11th January and 11th June 2024 for 10-minute resolution using our four case study locations. The raw data provided by Met Éireann is given in black and the model fit along with the 95\% credible interval is given in blue. The x-axis is the solar irradiance recorded at an 10-minute temporal resolution. }
    \label{fig:mod_fit_4_10min}
\end{figure}

As noted earlier, we use four case study locations, for one day in January and one day in June to highlight our model performance. Figure \ref{fig:mod_fit_4_10min} presents our 10-minute resolution model predictions along with the original meteorological station data. As expected, June exhibits a wider range of solar irradiance values and more pronounced short-term variations. It is clear that the 10-minute resolution data is successfully capturing many of the solar irradiance fluctuations within the 95\% credible interval which is shaded in light blue.  

\begin{figure}[H]
    \centering
    \begin{subfigure}[b]{0.48\linewidth}
        \centering
        \includegraphics[width=\linewidth]{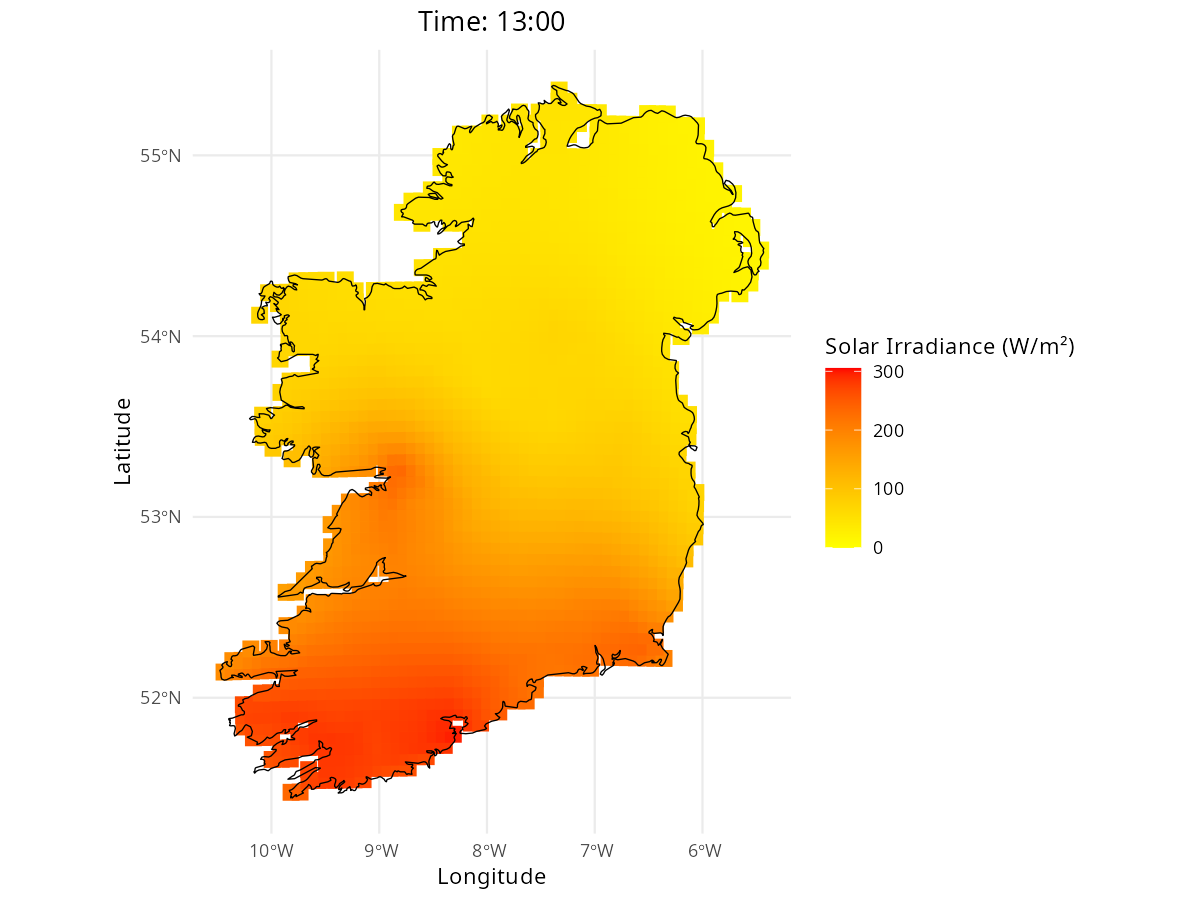}
        \caption{Posterior mean predicted solar irradiance for Ireland in W/m$^2$ on 11th January 2024 at 1pm}
        \label{fig:map_10min_jan_1pm}
    \end{subfigure}
    \hfill
    \begin{subfigure}[b]{0.48\linewidth}
        \centering
        \includegraphics[width=\linewidth]{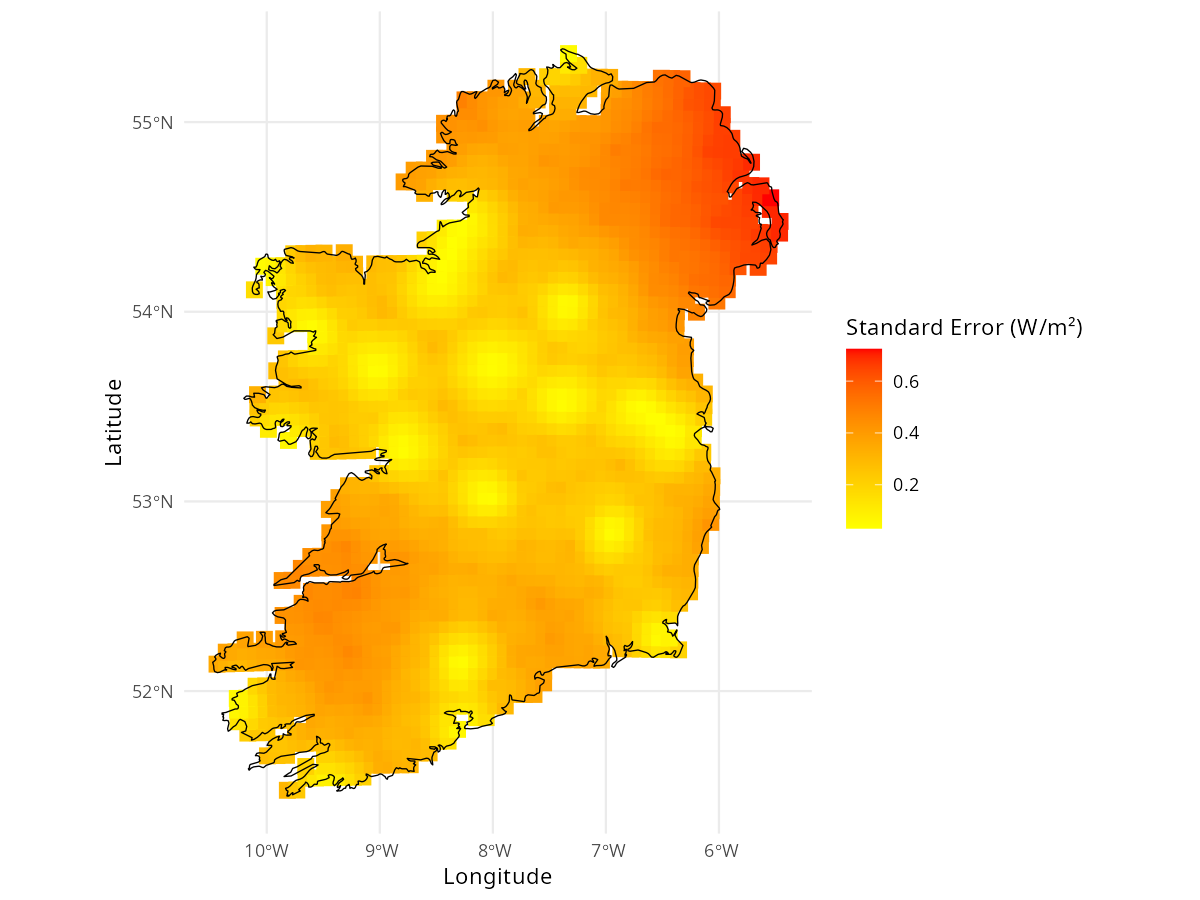}
        \caption{Standard error for posterior mean predicted solar irradiance for Ireland in W/m$^2$ on 11th January 2024 at 1pm}
        \label{fig:sd_map_10min_jan_1pm}
    \end{subfigure}
    \begin{subfigure}[b]{0.48\linewidth}
        \centering
        \includegraphics[width=\linewidth]{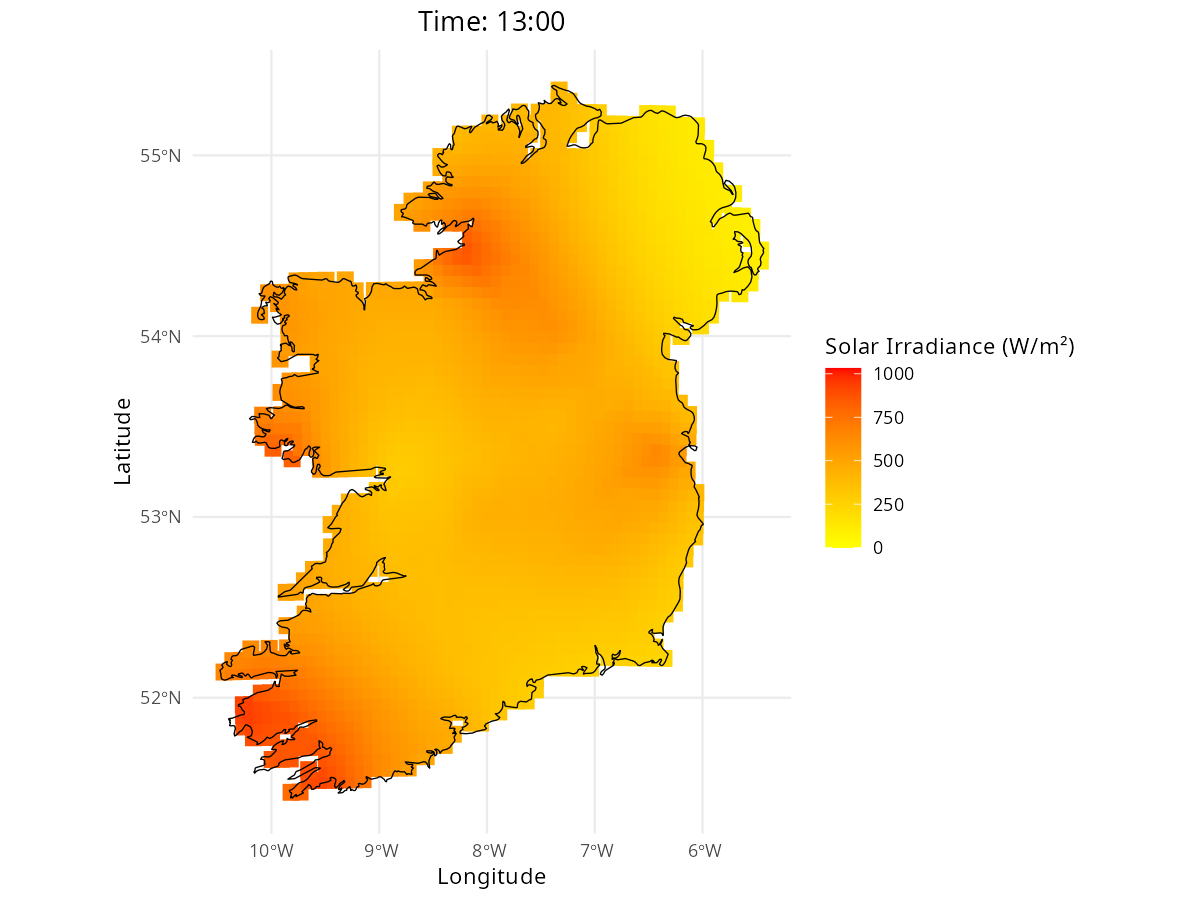}
        \caption{Posterior mean predicted solar irradiance for Ireland in W/m$^2$ on 11th June 2024 at 1pm}
        \label{fig:map_10min_jun_1pm}
    \end{subfigure}
    \hfill
    \begin{subfigure}[b]{0.48\linewidth}
        \centering
        \includegraphics[width=\linewidth]{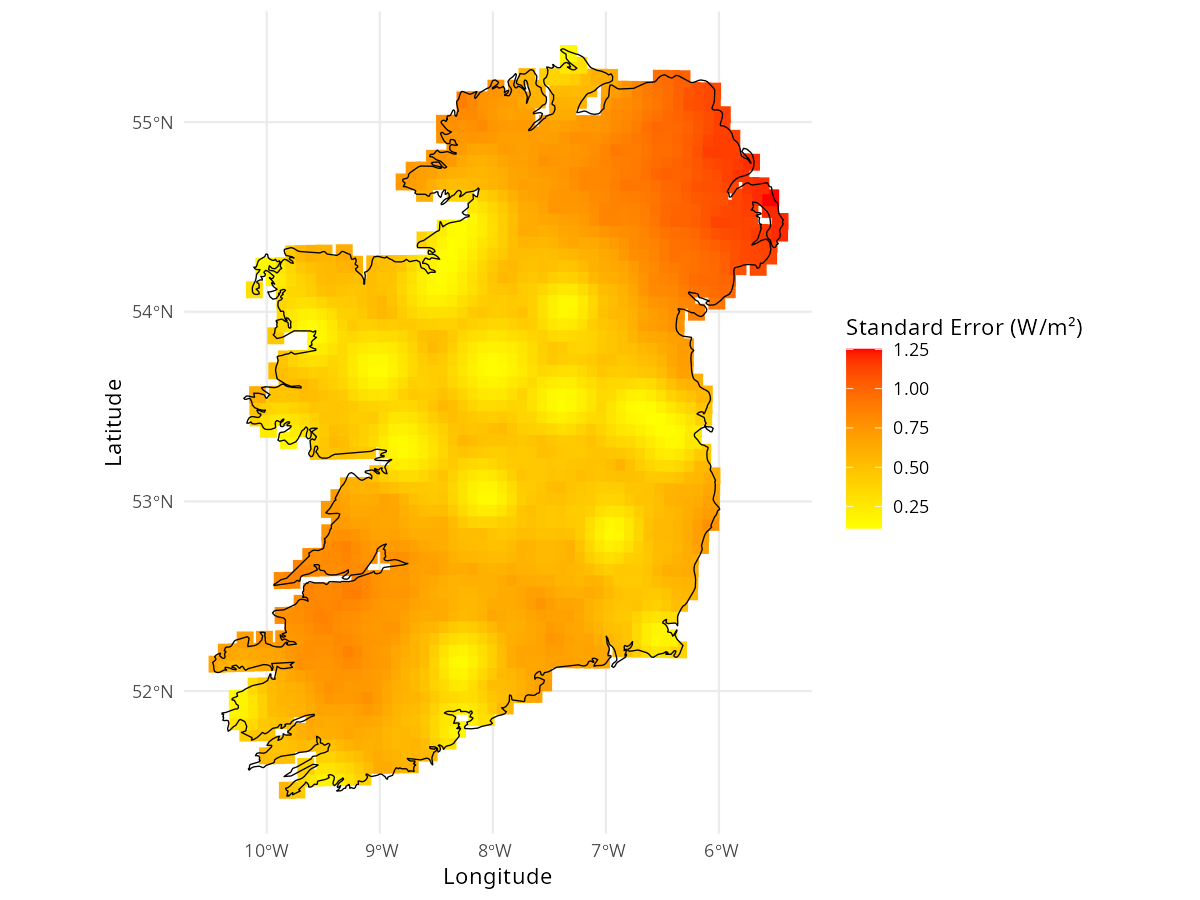}
        \caption{Standard error for posterior mean predicted solar irradiance for Ireland in W/m$^2$ on 11th January 2024 at 1pm}
        \label{fig:sd_map_10min_jun_1pm}
    \end{subfigure}

    \vspace{0.5cm}
    \caption{Posterior mean solar irradiance and associated standard errors across Ireland at 13:00 (1 pm) as estimated by a spatio-temporal Bayesian INLA model for January and June 2024 for 10-minute resolution. Each row corresponds to a month, with the left panels (a,c) showing the predicted solar irradiance surface (W/m²) and the right panel (b,d) showing the corresponding standard errors. The black outlines indicate the border of Ireland and Northern Ireland, highlighting spatial variation and model uncertainty for 10-minute resolution solar irradiance.}
    \label{fig:surf_hour_all_10}
\end{figure}

Furthermore, we can produce spatio-temporal maps of solar irradiance Ireland at 10-minute resolution. Figure \ref{fig:map_10min_jan_1pm} and Figure \ref{fig:map_10min_jun_1pm} present the estimated solar irradiance surface for Ireland for 10-minute resolution at 1 pm for the 11th of January and 11th of June 2024. It is clear that the maps have additional granularity at the 10-minute resolution compared with the hourly resolution. This is extremely important when predicting solar irradiance in different locations in Ireland which experience vastly different weather conditions. Figures \ref{fig:sd_map_10min_jan_1pm} and \ref{fig:sd_map_10min_jun_1pm} demonstrate the associated standard errors for both surfaces. There is slightly higher standard error at a 10-minute resolution compare with hourly, this is due to the increased fluctuations using this temporal resolution. As mentioned previously, larger standard errors are present in Northern Ireland reflecting the lack of observations in that area. 

\subsection{Model Validations}\label{mod_val_section}

To validate our model structure, we conducted a leave-one-station-out (LOSO) analysis for each temporal resolution, in which data from a single station is excluded and the model is trained on the remaining sites. Predictions are then generated for the excluded station, following the approach recommended by \citet{held2009posterior}. This process is repeated for every station in the dataset to ensure consistency. We evaluate the model performance, as shown in Table \ref{tab:mod_val}, based on the out-of-sample empirical coverage, mean absolute error (MAE), mean absolute percentage error (MAPE), mean bias error (MBE) and the Root Mean Squared Error (RMSE) as recommended by \citet{ZHANG2015157}. The empirical coverage provides the percentage of occasions that the true observation is within the model prediction interval (PI), which is a useful metric for investigating how well our solar irradiance estimates capture uncertainty and whether the prediction intervals are well-calibrated. The mean absolute error and the mean absolute percentage error can be used to evaluate uniform prediction errors, while the mean bias error is used to assess prediction bias \citep{ZHANG2015157}. The RMSE is used to evaluate prediction performance in the same units as the response (W/m$^2$).

Table \ref{tab:mod_val} provides the error metrics for the full dataset for 2024 for each temporal resolution. It is evident that the prediction accuracy decreases as the temporal resolution increases, this is due to the higher frequency variability of solar irradiance at finer time scales which is challenging to model and is not captured with a smooth spatio-temporal model. The hourly model achieves the lowest error( RMSE = 60W/$m^2$, MAE = 26W/$m^2$ and MBE which is close to zero), yet is overconfident as evident from the 95\% empirical coverage being 80\%, demonstrating that the prediction intervals are too narrow and underestimating the true variability. For the sub-hourly resolutions, the errors increase (RMSE up to 125 W/$m^2$) and the bias becomes stronger (MBE = -39.57 W/$m^2$), nevertheless the empirical coverage (95\% for 10-minute) demonstrates a better balance between precision and reliability highlighting their ability to quantify uncertainty better.
\begin{table}[h!]
\centering
 \begin{tabular}{|| p{2cm} | p{2cm} | p{2cm} | p{2cm} | p{2cm}| p{2cm}||} 
 \hline
Model Type & RMSE (W/m$^2$) & MAE (W/m$^2$) & MAPE (\%) & MBE (W/m$^2$) & 95 \% Empirical Coverage \\ [0.5ex] 
 \hline\hline
 Hourly & 60.05 & 26.02 &62.49 & -4.36 & 80\%\\ 
 \hline
 10-minute Model & 124.83 & 52.27 &70.03 & -39.57 & 95\% 
 \\[1ex] 
 \hline
 \end{tabular}
 \caption{Validation summary for the spatio-temporal model of solar irradiance at hourly and 10-minute resolutions. Leave-one-site-out cross-validation over one year was used to compute metrics—Root Mean Square Error (RMSE), Mean Absolute Error (MAE), Mean Absolute Percentage Error (MAPE), Mean Bias Error (MBE), and 95\% empirical coverage.}
\label{tab:mod_val}
\end{table}
In Table \ref{tab:mod_val_4sites}, we present the site-specific results for only the four case study sites for each temporal resolution with the remaining sites are shown in Appendix \ref{appendix_more_mod_res}. It is evident that Malin Head consistently possesses the largest errors, this is not surprising as it is a location on the northernmost part of the Irish coastline and is subject to extreme weather conditions \citep{brien2017extreme}. Gurteen and Phoenix Park generally have the lowest errors. For each location, the model is systematically over-predicting the solar irradiance for hourly and 10-minute resolutions (MBE: 23.43 to 33.61 W/$m^2$ and 48.29 to 106.68 W/$m^2$ respectively).
\begin{table}[ht]
\centering
\begin{tabular}{lrrrrr}
  \hline
  \multicolumn{5}{c}{\textbf{Hourly resolution}} \\ 
  \hline
Station & RMSE & MAE & MAPE & MBE & 95\% Empirical Coverage \\ 
  \hline
MALIN HEAD & 78.74 & 149.66 & -2.15 & 33.61 & 91\% \\ 
  MARKREE & 54.87 & 68.53 & -3.87 & 23.53 & 76\% \\ 
  PHOENIX PARK & 52.82 & 60.87 & 1.77 & 23.87 & 78\% \\ 
  GURTEEN & 51.58 & 44.38 & -7.23 & 23.43 & 78\% \\

 \hline
  \multicolumn{5}{c}{\textbf{10-minute resolution}} \\ 
  \hline
Station & RMSE & MAE & MAPE & MBE & 95\% Empirical Coverage  \\ 
  \hline
  MALIN HEAD & 176.80 & 83.61 & 136.38 & -80.09 & 85\% \\ 
MARKREE & 93.47 & 38.07 & 69.72 & -18.49 & 96\% \\
 PHOENIX PARK & 106.45 & 44.85 & 66.04 & -31.36 & 95\% \\
GURTEEN & 108.55 & 45.94 & 57.63 & -34.05 & 97\% \\ 
  \hline
\end{tabular}

\caption{Validation metrics for the spatio-temporal model of solar irradiance at hourly and 10-minute resolutions for each individual site. Leave-one-site-out cross-validation over one year was used to compute metrics—Root Mean Square Error (RMSE), Mean Absolute Error (MAE), Mean Absolute Percentage Error (MAPE), Mean Bias Error (MBE), and 95\% empirical coverage. Results are shown for four Irish meteorological stations \citep{met_eireann_web} as a case study; metrics for the remaining stations are provided in the Appendix \ref{appendix_more_mod_res}.}
\label{tab:mod_val_4sites}
\end{table}

\begin{figure}[H]
    \centering
    \includegraphics[width=\linewidth]{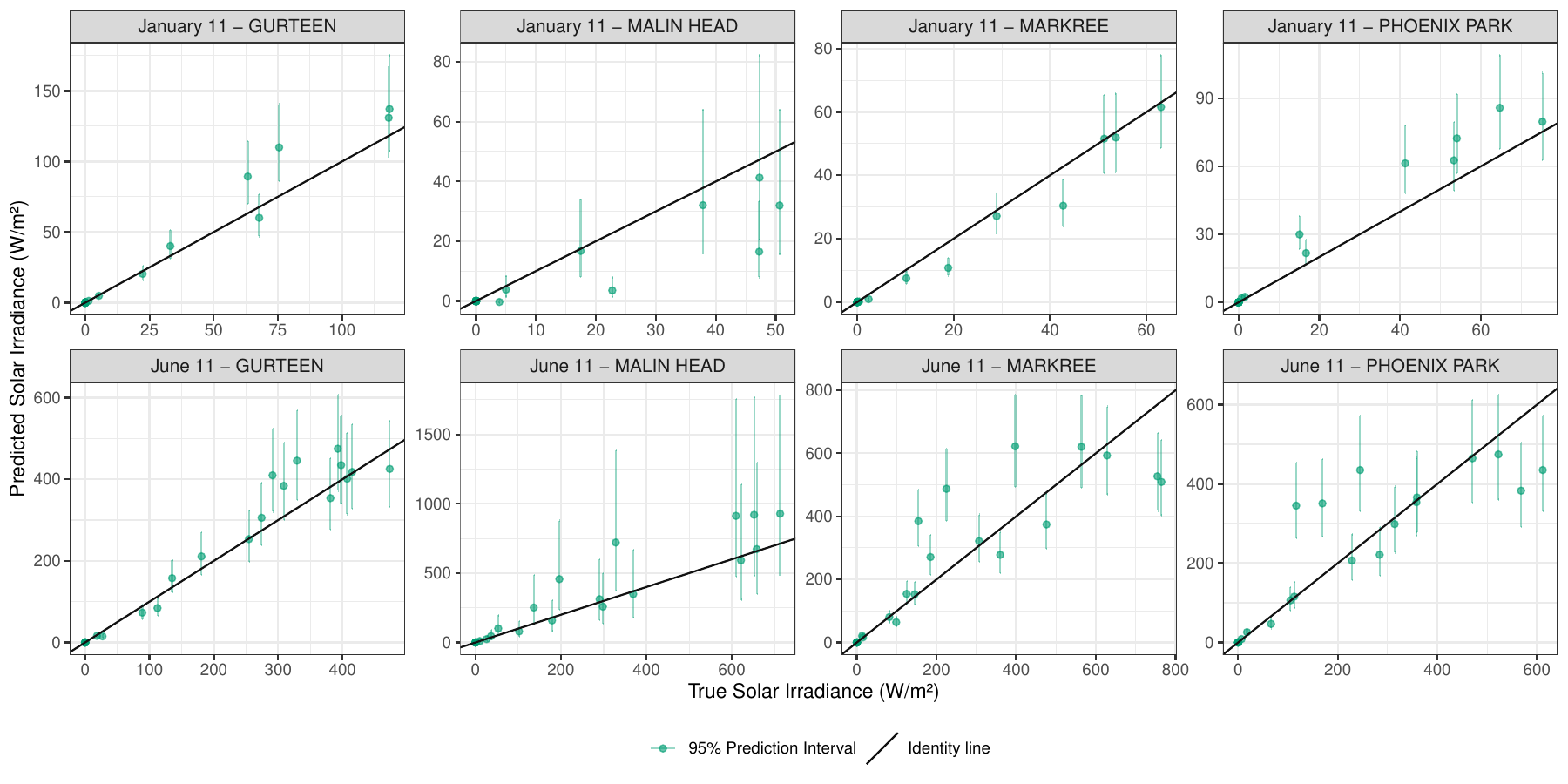}
    \caption{True versus predicted solar irradiance plots for hourly resolution for one day in January and June for four sites as a case study using the leave-one-site-out-cross validation checks. The 95\% prediction intervals are provided along with the identity line. }
    \label{fig:true_pred_hourly_4sites}
\end{figure}

\begin{figure}[H]
    \centering
    \includegraphics[width=\linewidth]{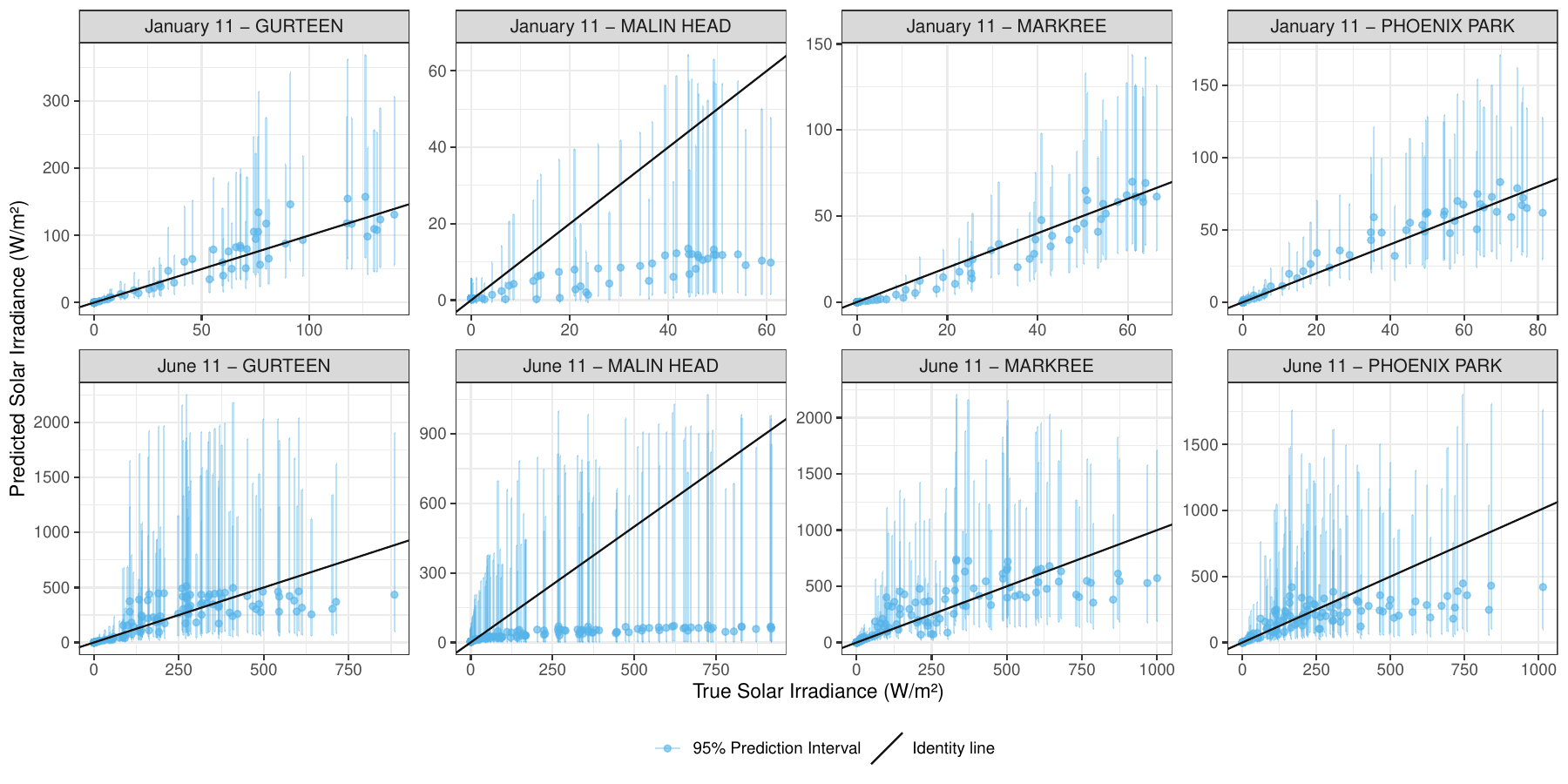}
    \caption{True versus predicted solar irradiance plots for 10-minute resolution for one day in January and June for four sites as a case study using the leave-one-site-out-cross validation checks. The 95\% prediction intervals are provided along with the identity line. }
    \label{fig:true_pred_10min_4sites}
\end{figure}

To further demonstrate our model performance, we visually demonstrate the model performance using true solar irradiance values plotted against the predicted solar irradiance values at an hourly (Figure \ref{fig:true_pred_hourly_4sites}) and sub-hourly resolutions (Figure \ref{fig:true_pred_10min_4sites}) for four case study locations for a day in January and June. It is clear that the hourly resolution model performs well for all sites, however June appears to possess more variability and larger prediction intervals for Malin Head. The 10-minute model shown in Figure \ref{fig:true_pred_10min_4sites} encounters difficulties for month of June and possess large 95\% prediction intervals to combat this but performs well in January.

\subsection{Comparing solar irradiance from different data sources}\label{other_datasource_test}
We have shown that our model performs well when assessed using cross-validation techniques, indicating that it is able to capture the underlying structure of solar irradiance variability. To further assess our model generalisability, we next validate the model using external data, providing a test of its accuracy and reliability on previously unseen conditions. In this section, we present comparisons for hourly and sub-hourly spatio-temporal models. 
\subsubsection{Hourly spatial temporal model}\label{era5_hourly}
To further validate our model, we compared the hourly spatio-temporal solar irradiance surface predictions with observations from local meteorological stations and the nearest grid points in the ERA5 reanalysis dataset \citep{era5_paper}. Since this reanalysis dataset is available only at hourly resolution, we used our hourly surface for a consistent comparison.

Table \ref{tab:mod_val_other_hour} presents the RMSE, MAE, MBE and MAPE values for the full year, comparing our hourly spatio-temporal model estimate for ground based measurements with ERA5 estimates for the same ground based measurements. This comparison highlights the differences between ground-based station measurements, the reanalysis datasets and our model, which provides predictions on the same spatial resolution as the reanalysis grids: $0.25^{\circ} \times 0.25^{\circ}$ for ERA5, covering the entire year of 2024. Our model achieved lower RMSE (35.85 vs. 82.16), MAE (15.35 vs. 44.05) and MAPE (35.85\% vs. 1449.86\%). Overall, the metrics in Table \ref{tab:mod_val_other_hour} demonstrates how our spatio-temporal model substantially outperforms ERA5 at the same spatial resolution across all major accuracy metrics highlighting the advantage of a tailored spatio-temporal statistical model over coarse global reanalysis product like ERA5. 

\begin{table}[!ht]
\centering
 \begin{tabular}{|| p{4cm} |p{2cm} |p{2cm} | p{2cm} | p{2cm} ||} 
 \hline
Type & RMSE & MBE & MAE & MAPE  \\ [0.5ex] 
 \hline\hline
  Hourly spatio-temporal model with same grid resolution as ERA5 & 35.85 & -3.28& 15.35& 35.85  \\ 
 \hline
 ERA5 & 82.16 & 1.41& 44.05& 1449.86 \\
 \hline
 \end{tabular}
 \caption{Validation metrics for our spatio-temporal model of hourly solar irradiance compared to reanalysis dataset, ERA5 \citep{era5_paper}. Metrics computed include Root Mean Square Error (RMSE), Mean Bias Error (MBE), Mean Absolute Error (MAE) and Mean Absolute Percentage Error (MAPE).}
\label{tab:mod_val_other_hour}
\end{table}

In Appendix \ref{appendix_era5}, we visually compare ERA5 estimates with the raw meteorological station data and our model predictions. For each measurement at a ground-based meteorological station location, we extracted the corresponding value from the closest ERA5 grid point and the model prediction at the same spatial resolution. Figures \ref{fig:mod_fit_jan_hour_era5_all_sites} and \ref{fig:mod_fit_jun_hour_era5_all_sites} illustrate how ERA5 aligns with the observed station data compared with our model, showing that our approach better captures site-specific variability that ERA5 fails to reproduce for a day in January and a day in June. Additionally, Figure \ref{fig:mod_fit_jun_hour_era5_all_sites} presents true versus predicted values for each station, with ERA5 plotted in red and our model in blue. Across all locations, the spread of variability is noticeably larger for ERA5 than for our model predictions, highlighting the improved accuracy and uncertainty representation of our approach.

\subsection{Investigating solar power output}\label{solar_power_output}
In this section, we use our spatio-temporal solar irradiance model at hourly and 10-minute resolutions as input to the \texttt{solaR} package to estimate solar power production. As described in Section \ref{solar_pv_data}, we validate our estimates using actual solar PV installation data. For the hourly model, we compare predicted power with residential solar PV data, ERA5 reanalysis data and observations from the closest meteorological station to assess the impact of different input data sources on solar production in Ireland. For the 10-minute model, we generate high-resolution power estimates and compare them with our two solar PV installation datasets at the same temporal resolution. To produce these estimates, we input the PV system characteristics into \texttt{solaR}, along with our spatio-temporal irradiance surfaces. 

\subsubsection{Hourly residential solar production analysis}
The first case study is our residential solar PV installation located in Dublin (Figure \ref{fig:map_data_sources}). We have hourly data covering the entire year of 2024 which enables us to generate hourly solar production estimates for the full year. 

\begin{figure}[ht]
    \centering
    \includegraphics[width=\linewidth]{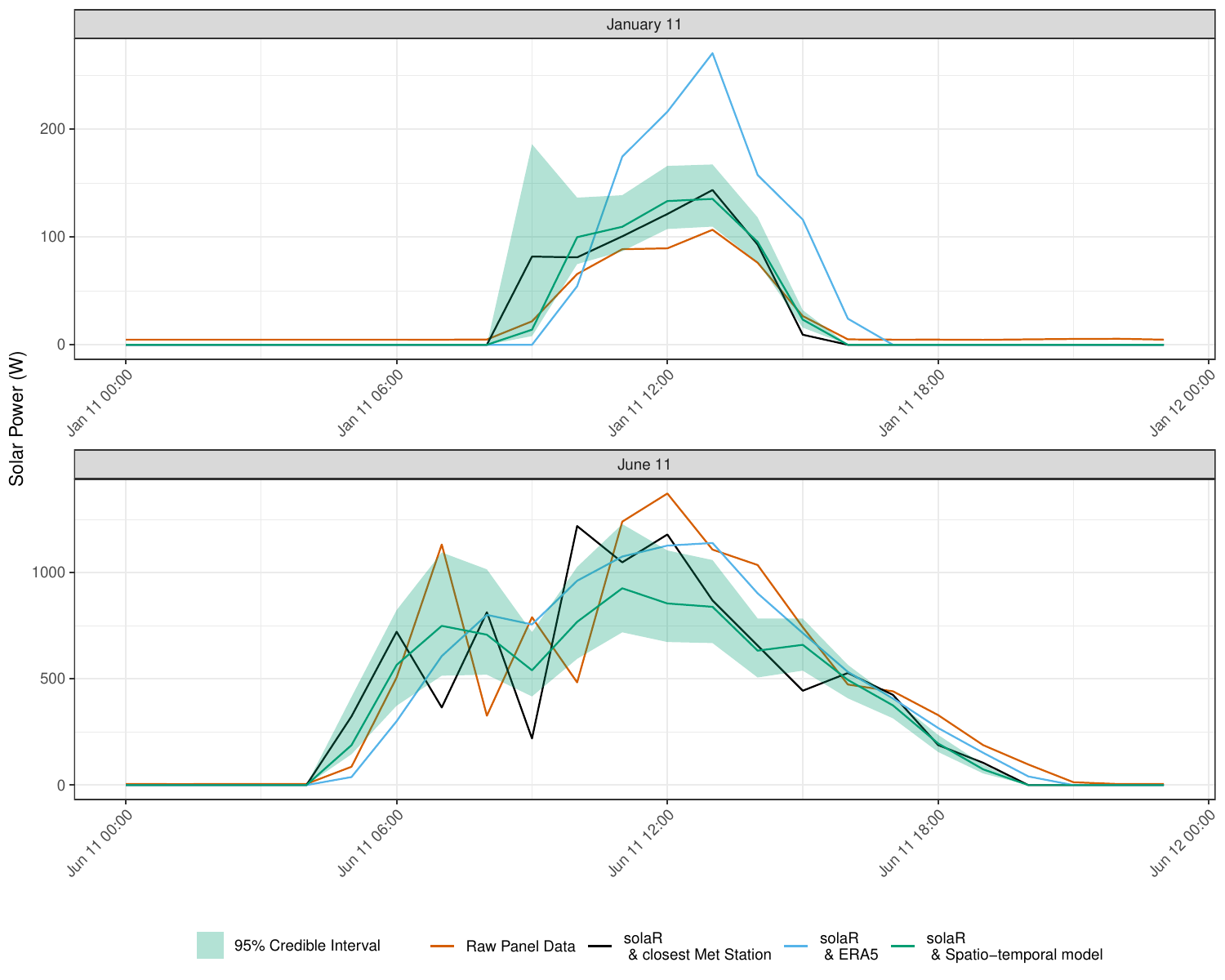}
    \caption{Comparison of observed and modelled solar power for a single day in winter (January 11) and summer (June 11) at the Dublin rooftop panel for hourly temporal resolution. Orange lines show the measured solar PV installation output, green lines show the spatio-temporal model predictions with the 95\% credible interval shaded in light green and coloured lines represent \texttt{solaR} predictions based on nearby meteorological stations (black) and ERA5 reanalysis (blue). Facets separate the two days to highlight seasonal differences in daily solar irradiance patterns.}
    \label{fig:hourly_panel_jun_jan_power}
\end{figure}

To illustrate how different input data sources impact solar production estimates, we plot the raw solar PV system data (orange line) with estimated solar power outputs using different input datasets for a day in June and January (Figure \ref{fig:hourly_panel_jun_jan_power}). As expected, output in January is much lower than in June, reflecting the seasonal differences in solar irradiance. Overall, all input datasets capture the observed patterns well, yet, our “solaR \& spatio-temporal model” (green line) uniquely provides a 95\% credible interval. This highlights that our modelling approach provides accurate estimates for solar power output along with uncertainty quantification which is an important advantage over the alternative approaches.

To further examine the impact of different input data sources into the \texttt{solaR} package and in turn, solar production estimates, we present Table \ref{tab:mod_val_solaR_hourly}. This table shows provides summary statistics for each input data source compared with the true solar PV output. It is clear that our spatio-temporal model produces solar power estimates comparable to ERA5 and the closest meteorological station. Our model achieves a low RMSE value of 225.83, indicating strong explanatory power, slightly higher then ERA5 (219.57) but lower than the closest Met Station approach (242.07). In terms of bias, the Mean Error (ME) is 44.87 W and the Mean Bias Error (MBE) is -44.87 W, showing that the model slightly underestimates solar power but less so than ERA5, which has larger negative errors (-50.66 W). Absolute errors are also low, with a Mean Absolute Error (MAE) of 108.25 W and an MAE relative to system size of 4.42\%, slightly outperforming ERA5 (108.99 W, 4.45\%) and the closest Met Station (114.72 W, 4.68\%). The Root Mean Square Error (RMSE) of 225.83 W further confirms that the model produces consistent predictions with moderate deviation from observed values. The Mean Absolute Percentage Error (MAPE) of 98.56\% reflects a reasonable relative accuracy for hourly predictions. Overall, the spatio-temporal model, at an hourly level, provides a balanced performance, reducing bias and absolute errors while maintaining comparable explanatory power to ERA5, and slightly improving upon the closest Met Station approach across most metrics. Importantly, our approach also allows for the incorporation of uncertainty, a feature not provided by the other two input data sources.

\begin{table}[ht]
\centering
\resizebox{\textwidth}{!}{%
\begin{tabular}{|| p{5cm} |  p{3cm} | p{3cm} | p{3cm} | p{3cm} | p{3cm} | p{3cm} ||} 
 \hline
 Model & ME [W] & MAE [W] & MAE relative to system size [\%] & RMSE [W] & MBE [W] & MAPE [\%] \\ 
 \hline\hline
 Closest Met Station \& \texttt{solaR} & 29.35 &  114.72 & 4.68 & 242.07 & -29.35 & 118.97 \\ 
 \hline
 ERA5 \& \texttt{solaR}  & 50.66 & 108.99 & 4.45 & 219.57 & -50.66 & 85.51 \\
 \hline
 Spatio-temporal model \& \texttt{solaR}  & 44.87 & 108.25 & 4.42 & 225.83 & -44.87 & 98.56 \\
 \hline
\end{tabular}%
}
\caption{Comparison of solar power prediction performance at hourly resolution (2024) for a residential PV installation using three different input data sources: closest meteorological station measurements, ERA5 reanalysis data, and a spatio-temporal model. Performance metrics include error measures (Mean Error, Mean Absolute Error, Root Mean Square Error, Mean Bias Error), relative error to system size and Mean Absolute Percentage Error (MAPE). This table allows direct comparison of model accuracy and bias across different evaluation criteria.}
\label{tab:mod_val_solaR_hourly}
\end{table}

\subsubsection{10-minute residential and industrial solar power analysis}\label{short_term_sol}
Examining sub-hourly solar power estimates are very useful for understanding short term fluctuations and managing their impact on the grid. With this in mind, we present our second case study which examines 10-minute solar power production using a residential and an industrial solar PV installation (Figure \ref{fig:map_data_sources}). For the residential and industrial case, we focus on two representative months, January and June 2024.

\begin{figure}
    \centering
    \includegraphics[width=\linewidth]{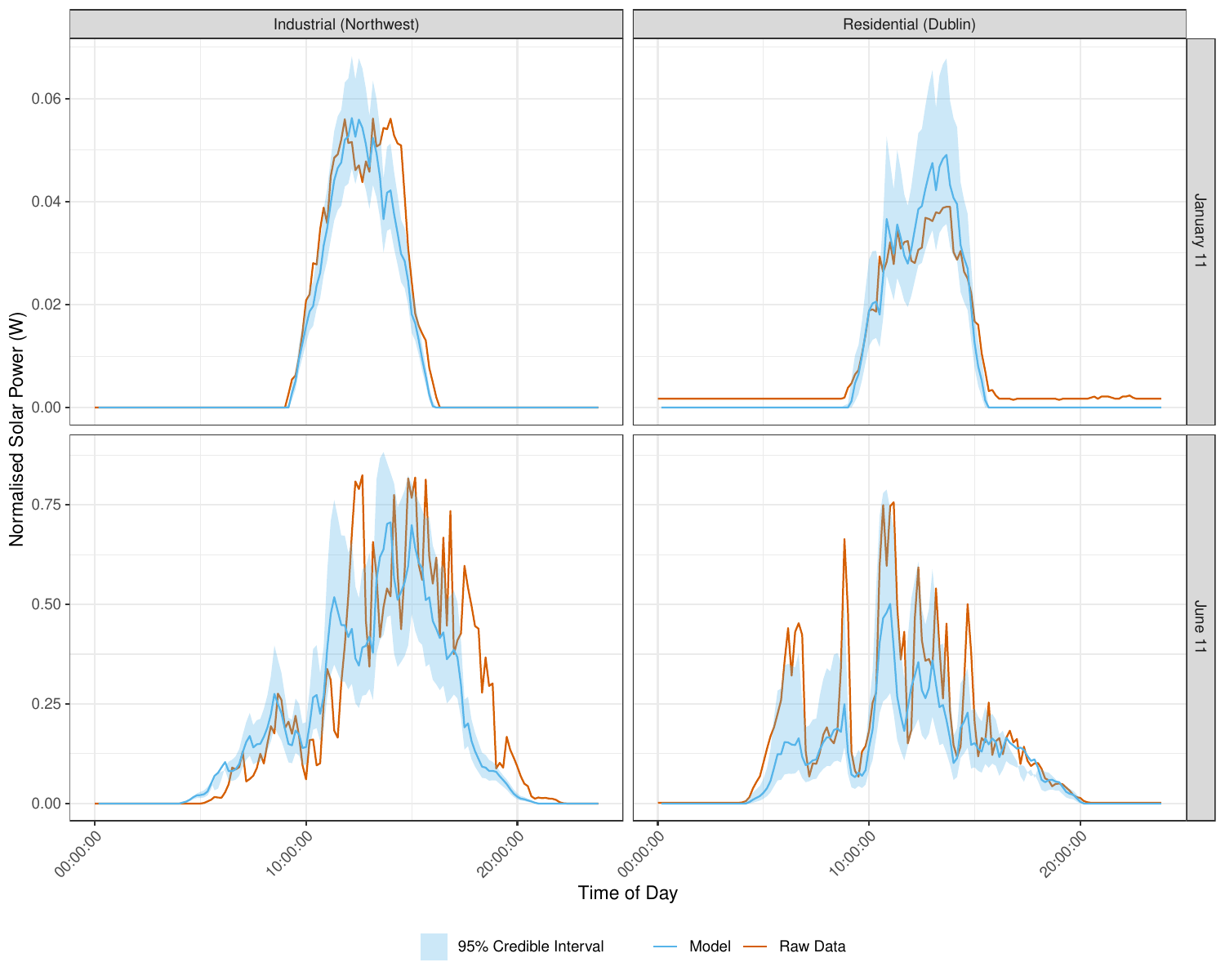}
    \caption{Comparison of observed and modelled solar power for a single day in winter (January 11) and summer (June 11) for an industrial site in Northwest (Industrial (Northwest)) and a residential site (Residential (Dublin)) in Dublin for 10-minute temporal resolution. The y-axis is the normalised solar power which is calculated by dividing the actual solar output by the system's potential maximum output and is used to ensure data privacy is maintained. Red lines show the measured solar PV installation output, blue lines show the spatio-temporal model predictions with the 95\% credible interval shaded in light blue. Facets separate the two days (January 11 and June 11) to highlight seasonal differences in daily solar irradiance patterns.}
    \label{fig:10min_panel_model_plot}
\end{figure}

In Figure \ref{fig:10min_panel_model_plot}, we visually compare the raw solar PV data with the predictions from our spatio-temporal model at 10-minute resolution. The figure shows the model fit for an industrial site in the Northwest of the country alongside a residential site in Dublin. To maintain data privacy, we plot normalised solar power, calculated as the actual solar output divided by the system’s potential maximum output rather than the inverter size, which could otherwise underestimate performance. It is important to note that January exhibits a much smaller range of output compared to June, with both systems reaching over 70\% of their potential solar power production.

To further assess our model performance at 10-minute resolution, we calculate a number of metrics shown in Table \ref{tab:mod_val_monthly_combined} for our industrial and residential sites in January and June. It is clear that our 10-minute spatio-temporal model performs consistently well across both winter and summer months for both our industrial and residential sites. As mentioned in Section \ref{datasets}, the industrial and residential sites difference significantly in magnitude and variability as evident in Table \ref{tab:mod_val_monthly_combined}. For the industrial site, errors are larger in June compared to January, with a mean error (ME) of 1696.5 W and 196.3W respectively, reflecting the higher generation of solar power in summer. The normalised error metrics, such as MAE \% and MAPE\% remain relatively stable indicating consistent relative model performance across months. In particular, the MAE relative to the system size \%, ranges from 1 - 7 \%, confirms that the model captures the daily production patterns proportionally to the system size. For the residential site, errors are smaller overall due to the lower solar production, with ME values of 70.5 W in January and 173.5 W in June. Again, the the MAE relative to the system size \%, ranging from 3 - 8\%, indicates that the model accurately captures the output of the residential solar system with high accuracy. Notably, both sites reach high fractions of their potential maximum output, with MAPE values showing that the model captures daily production trends effectively. Overall, the model performs well in both locations and across seasons, with larger absolute errors in the industrial site due to higher power output, but similar relative accuracy.

\begin{table}[ht!]
\centering
\resizebox{\textwidth}{!}{%
\begin{tabular}{|| l | r | r | l | r | r | r ||} 
\hline
\textbf{Industrial (Northwest)} & ME [W] & MAE [W] & MAE relative to system size \% & RMSE [W] & MBE [W] & MAPE \% \\ 
\hline
January & 196.3 & 704.5 & 1.8 & 2070.1 & -196.3 & 86.4 \\
June & 1696.5 & 3017.1 & 7.5 & 5222.6 & -1696.5 & 86.0 \\
\hline
\textbf{Residential (Dublin)} & ME [W] & MAE [W] & MAE relative to system size  \% & RMSE [W] & MBE [W] & MAPE \% \\ 
\hline
January & 70.5 & 73.9 & 3.0 & 201.9 & -70.5 & 85.9 \\
June & 173.5 & 194.6 & 7.9 & 359.0 & -173.5 & 61.8 \\
\hline
\end{tabular}%
}
\caption{Validation of solar power estimates for 10-minute resolution for January and June 2024 for industrial and residential sites using the spatio-temporal model. Performance metrics include error measures (Mean Error, Mean Absolute Error, Root Mean Square Error, Mean Bias Error), relative error to system size and Mean Absolute Percentage Error (MAPE).}
\label{tab:mod_val_monthly_combined}
\end{table}

Another motivation for developing sub-hourly solar irradiance estimates, and in turn solar power estimates at sub-hourly resolution, is to address overload clipping. As discussed in Section \ref{introduction} and highlighted by \citet{villoz2022model}, hourly PV simulations often underestimate clipping losses because they rely on averaged solar irradiance values. Clipping occurs when DC power exceeds the inverter’s AC capacity, with the surplus energy lost \citep{villoz2022model}. Averaging across an hour can hide short-term fluctuations that trigger clipping, leading to underestimation of losses and overestimation of annual generation—especially for high DC/AC ratios \citep[when the PV array is oversized relative to the inverter capacity, leading to more frequent power limiting:][]{KAEWNUKULTORN2024121402}. By operating at sub-hourly resolution (e.g., 10 minutes), our model captures these events more accurately without the computational burden of full minute-level simulations. 

\begin{figure}
    \centering
    \includegraphics[width=\linewidth]{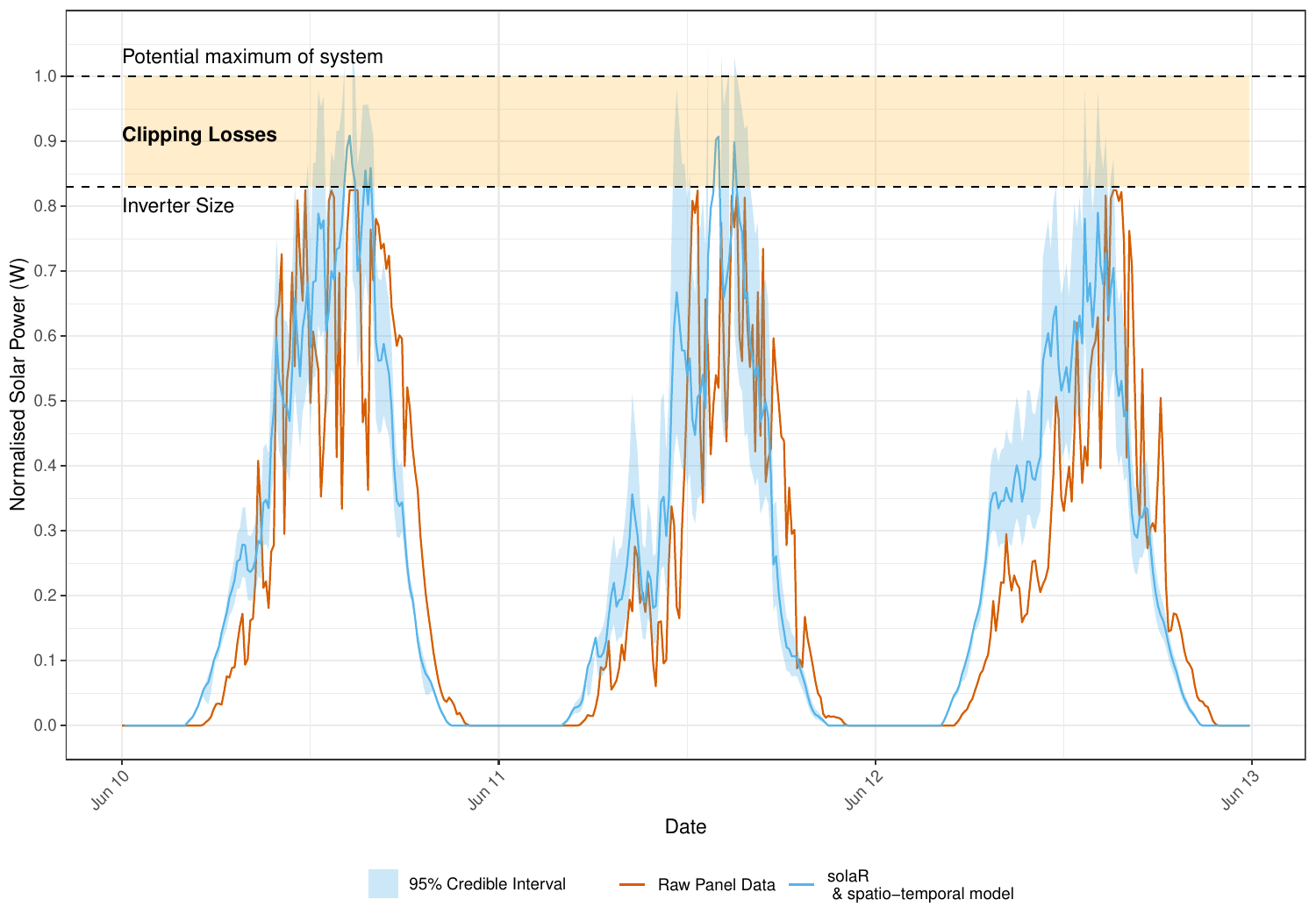}
    \caption{Three representative days of solar PV output in June 2024 for industrial site in the Northwest. The orange line shows the measured solar PV power at 10-minute resolution. The blue line with shaded band represent the modelled solar production and its 95\% credible interval. The lower black dashed line represents the inverter capacity and the upper black line represents the potential maximum of system. The shaded rectangle above the inverter size line illustrates the region of potential clipping losses, where predicted generation exceeds the inverter size. This highlights how short-duration peaks can be curtailed by the inverter and how the model identifies periods of overload clipping.}
    \label{fig:clip_week}
\end{figure}

For our industrial site, we applied the model to predict solar power generation without the inverter limit, allowing us to identify periods where the power would exceed the inverter size. From this, we can estimated the energy lost due to overload clipping. In Figure \ref{fig:clip_week}, we show three representative days of raw solar PV output (orange line) along with our modelled solar production (blue line) and the associated 95\% credible interval (blue shaded region). The lower horizontal line represents the inverter size and the upper horizontal line represents the potential maximum that the system could generate. The orange shaded region between the inverter limit and the system maximum highlights the potential clipping region. This area represents periods where the predicted power would exceed the inverter size, allowing us to visually identify overload clipping events. By comparing the raw PV output to the model predictions and the clipping region, the plot demonstrates how sub-hourly variations in solar irradiance can lead to short-term peaks that exceed the inverter capacity, which would be missed if only hourly data were considered. 

For the month of June, we estimated the losses due to clipping to be 1.05\% at a 10-minute resolution which aligns with previous research by \citet{villoz2022model}. In contrast, when the same calculations are performed using hourly-aggregated data, the estimated losses are slightly lower, at 0.83\%. This difference arises because hourly resolution smooths out short-duration peaks, which can temporarily exceed the inverter capacity. As a result, hourly resolution tends to underestimate the true energy losses due to clipping and highlights the advantage of using sub-hourly approaches. 

\section{Discussion}

In this paper, we introduced a novel Bayesian spatio-temporal statistical model to investigate hourly and sub-hourly solar irradiance across Ireland, while accounting for uncertainties in the underlying data. We utilised high temporal resolution open-source data from meteorological stations \citep{met_eireann_web} and provided solar irradiance predictions at any Irish location and historic time point. The model demonstrated strong performance in capturing both spatial and temporal variability in solar irradiance and performed comparatively well with external data sources such as reanalysis data.

We demonstrated the robustness of our statistical model structure using cross validation techniques in Section \ref{mod_val_section}. The leave-one-site-out analysis demonstrated that modelling at an hourly resolution provided the highest prediction accuracy, however, the empirical coverage, which gives insight into the reliability of the model’s uncertainty quantification, was lower than expected. This highlights how aggregating solar irradiance to hourly resolution underestimates true variability of the underlying response and while modelling at sub-hourly resolution introduces higher prediction error, it better reflects the fluctuations that are critical for solar production modelling. Additionally, we validated our approach using reanalysis data from ERA5 \citep{era5_paper}, as described in Section \ref{other_datasource_test}. As ERA5 data is limited to hourly resolution and is not available in real-time \citep{era5_paper}, it serves primarily as a benchmark. At the hourly resolution, our model outperformed ERA5 reanalysis data with more accurate predictions and reduced variability, as shown in Table \ref{tab:mod_val_other_hour}. These result highlight that our approach not only provides improved accuracy compared with ERA5, but also provides explicit quantification of uncertainty and allows for near real-time solar predictions.

We further demonstrated the applicability of our spatio-temporal irradiance model by using it as input for the open-source PV modelling package \texttt{solaR} to estimate residential solar power output. The results shown in Section \ref{solar_power_output}, demonstrates that our model provides competitive or improved accuracy compared to ERA5 and closest-station inputs at hourly resolution, particularly in terms of bias reduction and absolute error. By extending our model to 10-minute resolution and developing solar irradiance maps across Ireland  (e.g., Figure \ref{fig:map_10min_jun_1pm}), we provide a high resolution view of both temporal and spatial variability in solar resources. These maps are key for more detailed future analysis, such as understanding the implications short-term fluctuations in solar power production crucial for planning and managing the Irish electricity grid \citep{maimo2025wind}. With sufficient data on distributed Irish solar PV installations, this framework could be extended to evaluate how the growing penetration of rooftop solar alters electrical system behaviour across regions. Such an extension would allow quantification of its potential impact on grid stability and management, which is particularly relevant given that “small-scale embedded solar is not included” in national grid reporting \citep{smartgriddashboard2025}.

In our case studies in Section \ref{short_term_sol}, we highlighted that for both residential and industrial sites, our model can provide accurate estimates of solar production along with uncertainty, thereby offering not only point predictions but also measures of confidence to inform decisions. The sub-hourly modelling framework enabled us to quantify the impact of overload clipping, an increasingly important challenge as solar PV capacity expands and inverter limitations are more frequently reached \citep{KAEWNUKULTORN2024121402}. By capturing this effect, our approach offers additional insights into system performance under real operating conditions and highlights the growing need to account for clipping losses in grid integration studies.

There are a number of potential extensions for our statistical model. Firstly, our flexible model structure could allow for the inclusion of additional covariates, for example wind speed, which could allow for joint modelling of renewable energy generation. Issues arise when predicting solar irradiance for new location where these covariates do not have values, yet work by \citet{chacon2024mapping} described a new method of predicting this unobserved latent factor in new locations which could be used in future applications. However, due to the spatial nature of our variables, additional checks would be required to ensure spatial confounding is avoided \citep[e.g.][]{dupont2022spatial}.

Our approach uses a Bayesian framework which provides flexible model structures along with full uncertainty profiles for unknown parameters. Another benefit of a Bayesian framework is the ability to incorporate different data sources to help inform model estimates. A potential future direction would be to incorporate data from reanalysis datasets or physical models to help inform key covariates, similar to work in Spain by \citep{BEGUERIA2026123943}, and apply it to the highly variable Irish climate \citep{https://doi.org/10.1002/met.1892}. An alternative approach to incorporating external data sources is the use of satellite imaging in combination with traditional methods for predicting solar radiation, as demonstrated by \citet{Attya25}. However, both methods comes with challenges as resolution of these additional data sources vary both spatially and temporally. Other studies have shown the negative impact of aggregation when solar forecasting \citep{rs17060998}. Similarly, spatial aggregation can impact solar forecasting especially when working on large spatial scales or when data biases appear in the data at global scales \citep{brinkerink2024role}. Recent research has highlighted the potential of using citizen science data within a Bayesian framework to improve wind speed predictions which could be implemented for solar irradiance to overcome spatial sparsity \citep{organ2025enhancing}.

In order to improve the spatial spread of solar measurements in Ireland, we could incorporate solar panel data as an additional data source. This requires an extra step to convert the solar power observed from solar panels into the solar irradiance measurements recorded by weather stations. This conversion process, known as reverse transposition, has been studied by \citet{BERTRAND2018306}; however, the method encountered difficulties when converting data for certain solar positions. More recently, \citet{DRIESSE2024112093} proposed an improved reverse transposition approach. Nonetheless, limitations remain, particularly in assigning unique values of global horizontal irradiance when the angle of incidence approaches or exceeds 90$^{\circ}$.

Our approach has successfully delivered solar irradiance predictions at both hourly and sub-hourly resolutions, as well as solar power predictions for historical time points and in near real-time across any location in Ireland. An extension of this work is to investigate the potential of these techniques for solar forecasting. Previous studies highlight that machine learning algorithms are particularly effective for short-term forecasting, with hybrid approaches combining statistical and machine learning methods showing strong potential (e.g., \citet{wang2019echo} for wind power forecasting). 

Another important future work is to quantify national-scale electricity generation scenarios under varying levels of solar panel deployment, similar to the community-based microgeneration studies of \citet{en12234600}. Our statistical framework provides a foundation for probabilistic, countrywide estimates of solar energy generation under increasing adoption scenarios. Such insights would be relevant for addressing grid management challenges and informing long-term grid development strategies.

\section{Declaration of generative AI and AI-assisted technologies in the writing process}

During the preparation of this work the author(s) used GPT4 of openAI in order to improve the language and structure. After using this tool/service, the author(s) reviewed and edited the content as needed and take(s) full responsibility for the content of the publication.

\section{Acknowledgements}
This work is funded by Research Ireland under the EU Commission Recovery and Resilience Facility under the Research Ireland Energy Innovation Challenge Grant Number 22/NCF/EI/11162G. Organ's work is funded by Taighde Éireann – Research Ireland under Grant number 18/CRT/6049.

We would like to acknowledgement the support of Met Éireann in regard to data queries. We acknowledge the support of Dr Oscar Perpiñán Lamigueiro regarding software information. We would like to thank UrbanVolt for generously providing data in this study. 

\section{Appendix}

\subsection{Weather Station Data}
\begin{sidewaysfigure}
\begin{figure}[H]
    \centering
    \includegraphics[width=1\linewidth]{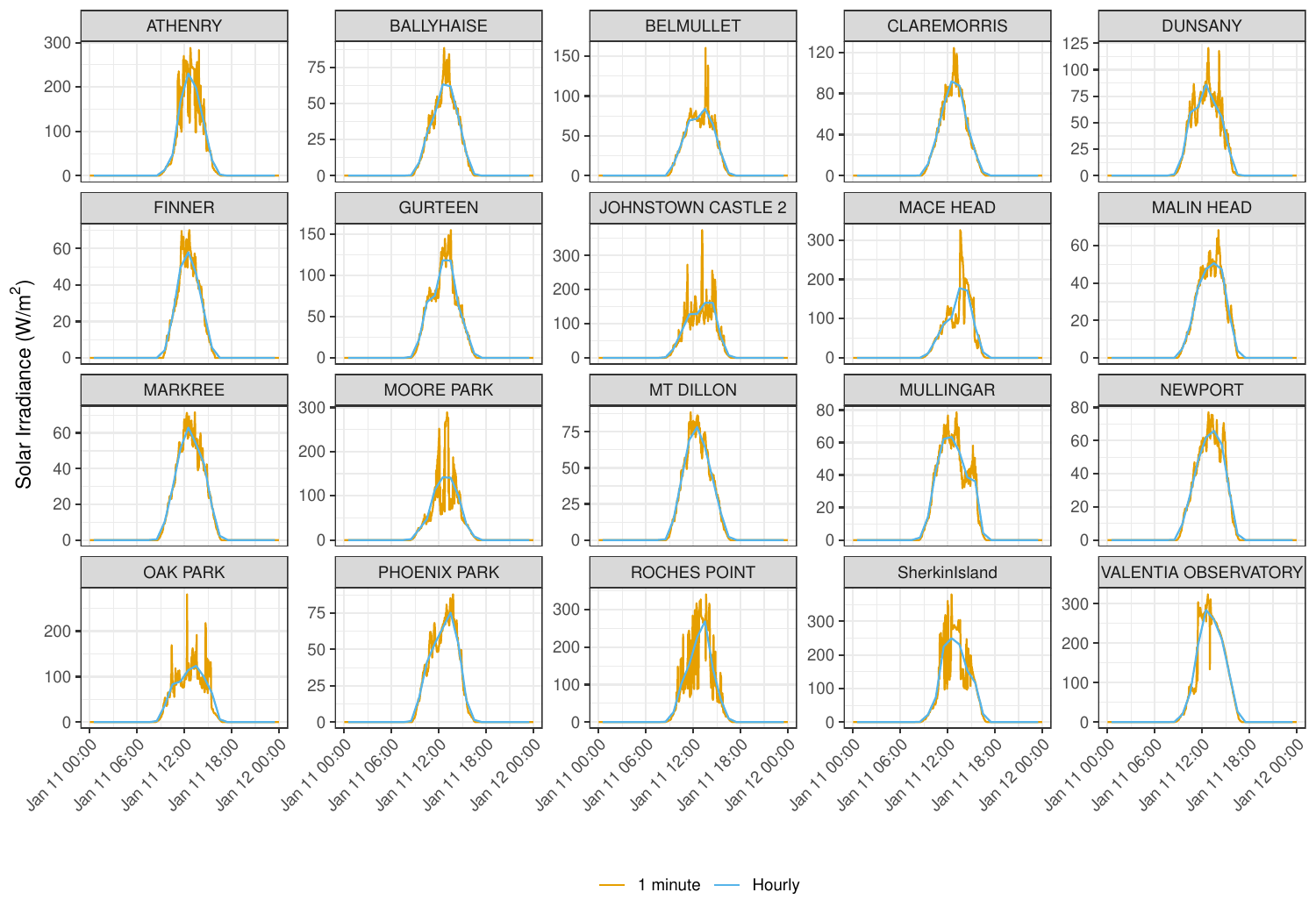}
    \caption{Plot of solar irradiance at 1 minute resolution from all weather stations for 11th January 2024 \citep[data provided by Met Éireann][]{met_eireann_web}. Certain sites experience more variability then others, for example Valentia Observatory and Phoenix Park.}
    \label{fig:raw_1min_all_sites_jan}
\end{figure}
\end{sidewaysfigure}

\begin{sidewaysfigure}
\begin{figure}[H]
    \centering
    \includegraphics[width=1\linewidth]{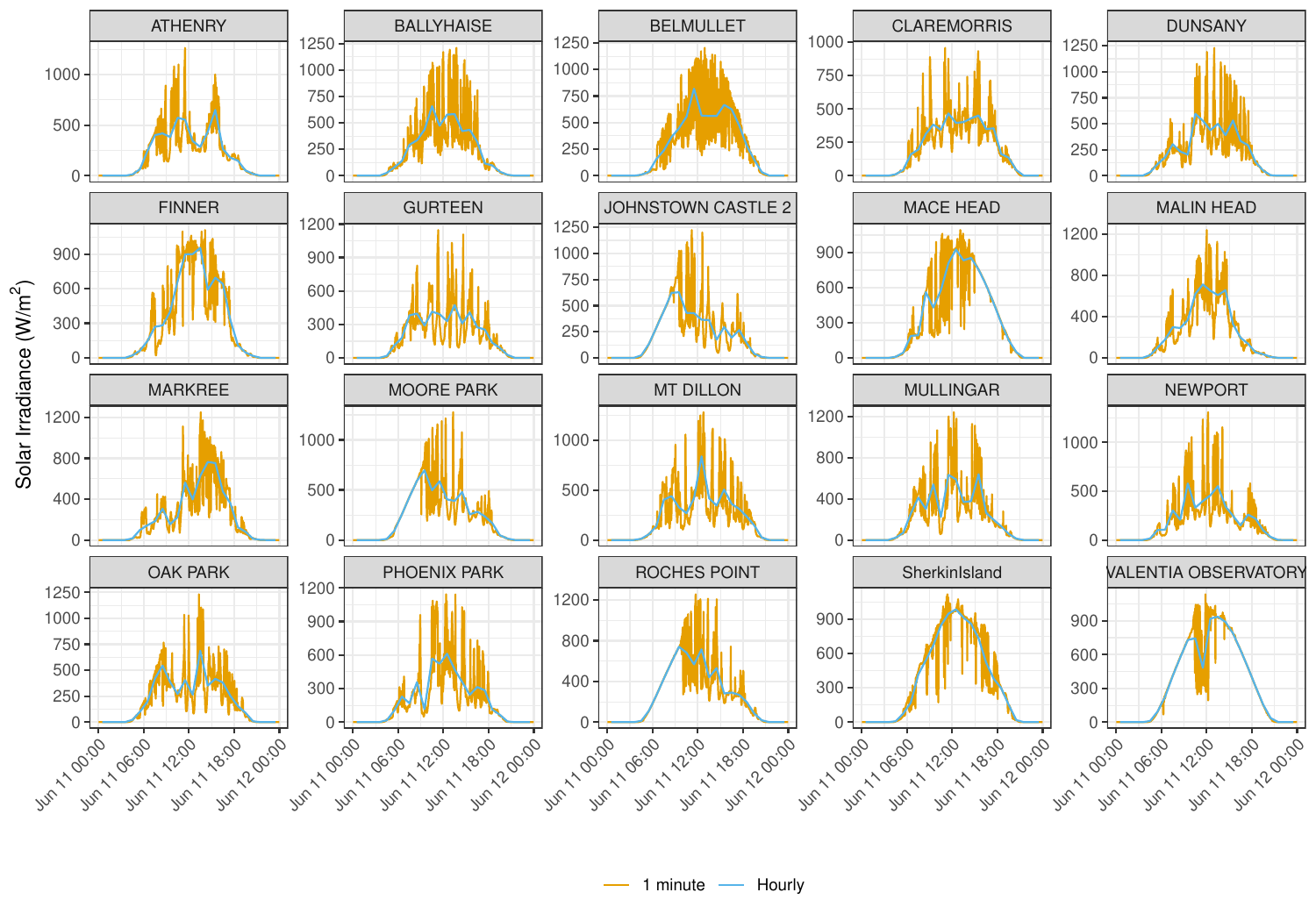}
    \caption{Plot of solar irradiance at 1 minute resolution from all weather stations for 11th June 2024 \citep[data provided by Met Éireann][]{met_eireann_web}. Certain sites experience more variability then others, for example Valentia Observatory and Malin Head.}
    \label{fig:raw_1min_all_sites_june}
\end{figure}
\end{sidewaysfigure}

For completeness, we have included the plots of 1 minute resolution data, supplied by Met Éireann \citep{met_eireann_web} for a day in January and a day in June shown in Figure \ref{fig:raw_1min_all_sites_jan} and Figure \ref{fig:raw_1min_all_sites_june}. These plots demonstrate the extreme variability of solar irradiance across the country, over the day and for different months of the year.

\newpage

\subsection{Additional Model Results}\label{appendix_more_mod_res}
In this section, we present results for complete dataset, i.e. the full set of stations. 
\subsubsection{Hourly Resolution}
\begin{sidewaysfigure}
\begin{figure}[H]
    \centering
    \includegraphics[width=\linewidth]{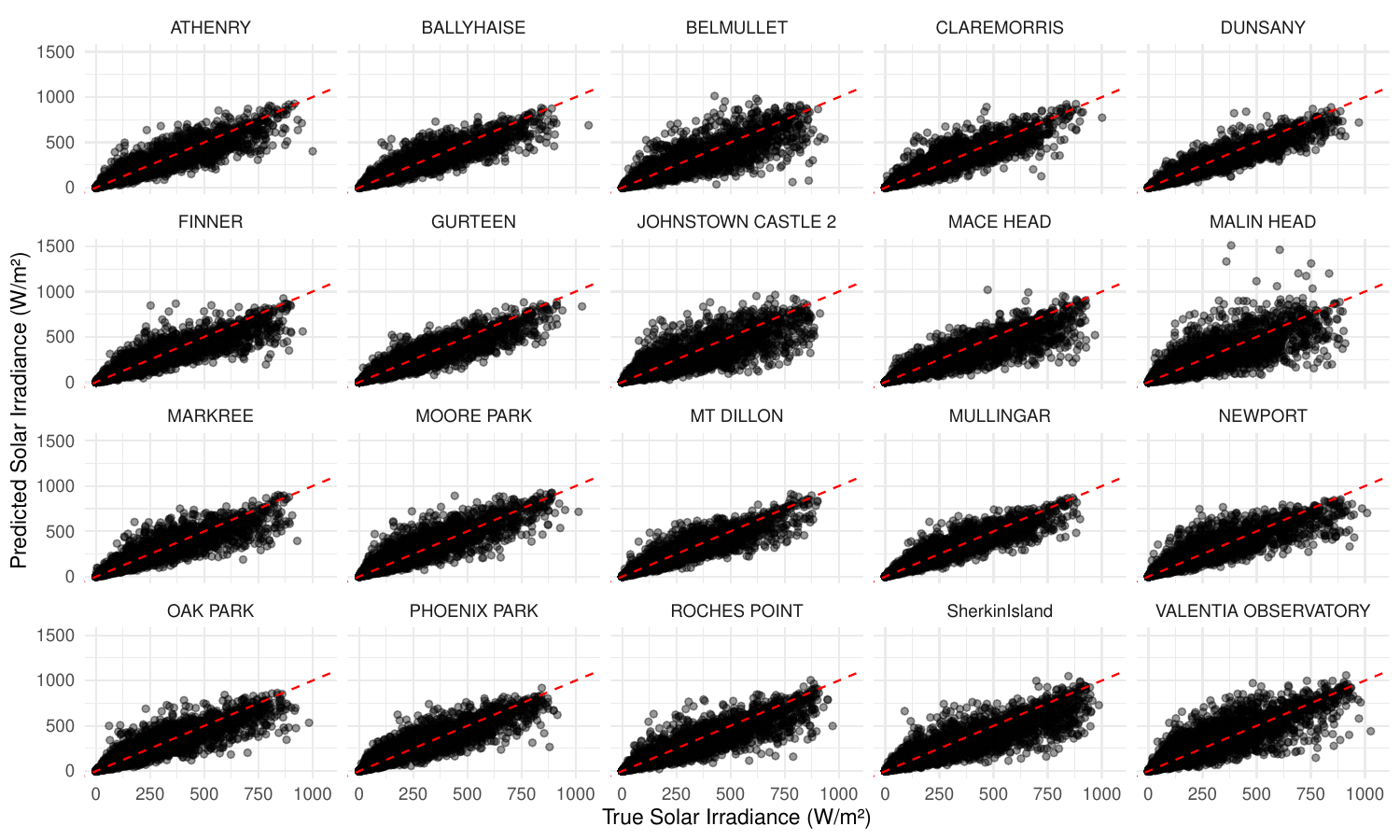}
    \caption{True versus predicted solar irradiance plot for each site for 1 full year at hourly resolution.}
    \label{fig:true_pred_1_year_hourly}
\end{figure}
\end{sidewaysfigure}

In this section, we present results from our hourly resolution spatio-temporal model across all 20 meteorological stations. Table \ref{tab:mod_val_full} reports summary statistics for each site, while Figure \ref{fig:true_pred_1_year_hourly} presents the true versus predicted plot for each location. 
\begin{table}[ht]
\centering
\begin{tabular}{lrrrrr}
  \hline
Station & RMSE & MAPE & MBE & MAE & 95\% Empirical Coverage \\ 
  \hline
MALIN HEAD & 78.74 & 149.66 & -2.15 & 33.61 & 0.91 \\ 
  SherkinIsland & 75.55 & 50.91 & -14.33 & 33.25 & 0.85 \\ 
  JOHNSTOWN CASTLE 2 & 73.81 & 59.44 & -8.14 & 32.69 & 0.83 \\ 
  VALENTIA OBSERVATORY & 73.15 & 78.35 & 6.29 & 32.44 & 0.84 \\ 
  MACE HEAD & 68.41 & 42.94 & -15.97 & 29.31 & 0.81 \\ 
  BELMULLET & 68.13 & 72.43 & -6.28 & 29.11 & 0.82 \\ 
  ROCHES POINT & 62.06 & 45.28 & -10.46 & 27.22 & 0.82 \\ 
  NEWPORT & 60.51 & 132.64 & 0.45 & 26.07 & 0.74 \\ 
  FINNER & 57.95 & 58.88 & -6.95 & 25.19 & 0.78 \\ 
  MOORE PARK & 56.90 & 74.13 & 5.83 & 25.75 & 0.78 \\ 
  OAK PARK & 56.51 & 43.19 & -3.85 & 25.08 & 0.78 \\ 
  MARKREE & 54.87 & 68.53 & -3.87 & 23.53 & 0.76 \\ 
  PHOENIX PARK & 52.82 & 60.87 & 1.77 & 23.87 & 0.78 \\ 
  ATHENRY & 51.60 & 57.15 & -4.11 & 22.83 & 0.78 \\ 
  GURTEEN & 51.58 & 44.38 & -7.23 & 23.43 & 0.78 \\ 
  CLAREMORRIS & 51.02 & 38.27 & -1.42 & 22.34 & 0.76 \\ 
  BALLYHAISE & 50.50 & 50.04 & -3.90 & 22.44 & 0.80 \\ 
  MT DILLON & 48.28 & 39.73 & -3.75 & 21.15 & 0.78 \\ 
  MULLINGAR & 46.03 & 47.36 & -0.97 & 20.44 & 0.77 \\ 
  DUNSANY & 45.50 & 37.88 & -8.23 & 20.65 & 0.77 \\ 
   \hline
\end{tabular}
\caption{Model validation for hourly solar irradiance using leave-one-site-out for one year. Leave-one-site-out cross-validation over one year was used to compute metrics—Root Mean
Square Error (RMSE), Mean Absolute Error (MAE), Mean Absolute Percentage Error (MAPE), Mean Bias Error
(MBE) and 95\% empirical coverage} 
\label{tab:mod_val_full}
\end{table}

\subsubsection{Compare with ERA5}\label{appendix_era5}
\begin{sidewaysfigure}
\begin{figure}[H]
    \centering
    \includegraphics[width=\linewidth]{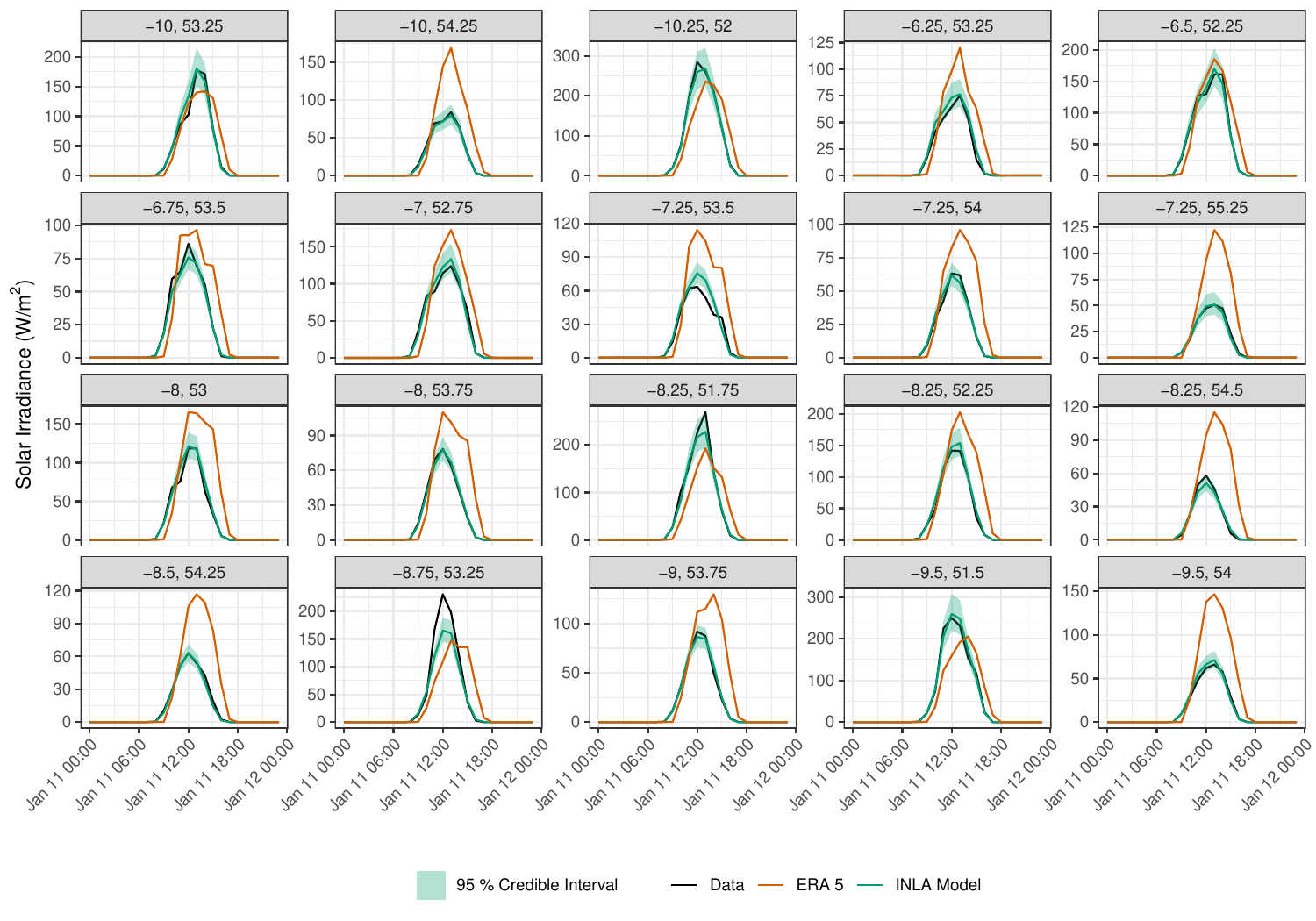}
    \caption{Station-level comparisons of solar irradiance from 20 Met Éireann ground-based stations (black), ERA5 estimates (orange), and spatio-temporal model predictions with 95\% credible intervals (green) on 11 January 2024.}
    \label{fig:mod_fit_jan_hour_era5_all_sites}
\end{figure}
\end{sidewaysfigure}

\begin{sidewaysfigure}
\begin{figure}[H]
    \centering
    \includegraphics[width=\linewidth]{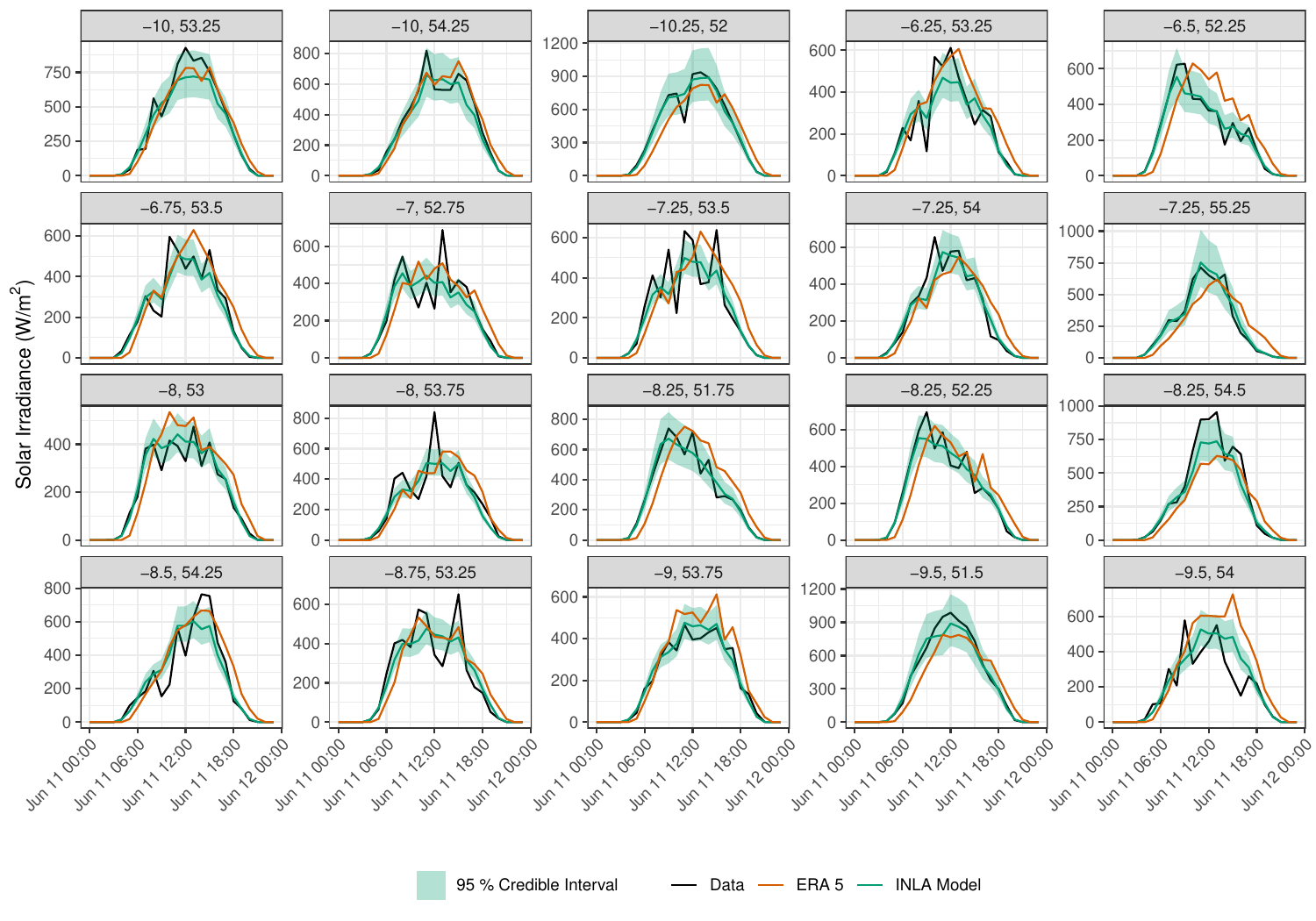}
    \caption{Station-level comparisons of solar irradiance from 20 Met Éireann ground-based stations (black), ERA5 estimates (orange), and spatio-temporal model predictions with 95\% credible intervals (green) on 11 January 2024.}
    \label{fig:mod_fit_jun_hour_era5_all_sites}
\end{figure}
\end{sidewaysfigure}

In Section \ref{era5_hourly}, we compared our hourly spatio-temporal model with the ERA5 reanalysis dataset. In this section, we provide detailed station-level comparisons between ground-based measurements from 20 meteorological station provided by \citet{met_eireann_web}, ERA5 estimates and our model predictions. For each meteorological station, we extracted the ERA5 grid-point value corresponding to its location and compared it with both the observed station data and our model predictions at the same spatial resolution. Figures \ref{fig:mod_fit_jan_hour_era5_all_sites} and \ref{fig:mod_fit_jun_hour_era5_all_sites} illustrate these comparisons for all 20 stations provided by Met Éireann \citep{met_eireann_web}, showing results for a representative day in January and a representative day in June. In addition, Figure \ref{fig:mod_fit_jun_hour_era5_all_sites} presents true versus predicted solar irradiance at each of the 20 meteorological stations, comparing our model predictions (blue) with ERA5 estimates (red). Both sets of predictions broadly align with the 1:1 identity line (black); however, our model predictions cluster more closely around the line, while ERA5 estimates show a noticeably larger spread.

\begin{sidewaysfigure}
\begin{figure}[H]
    \centering
    \includegraphics[width=\linewidth]{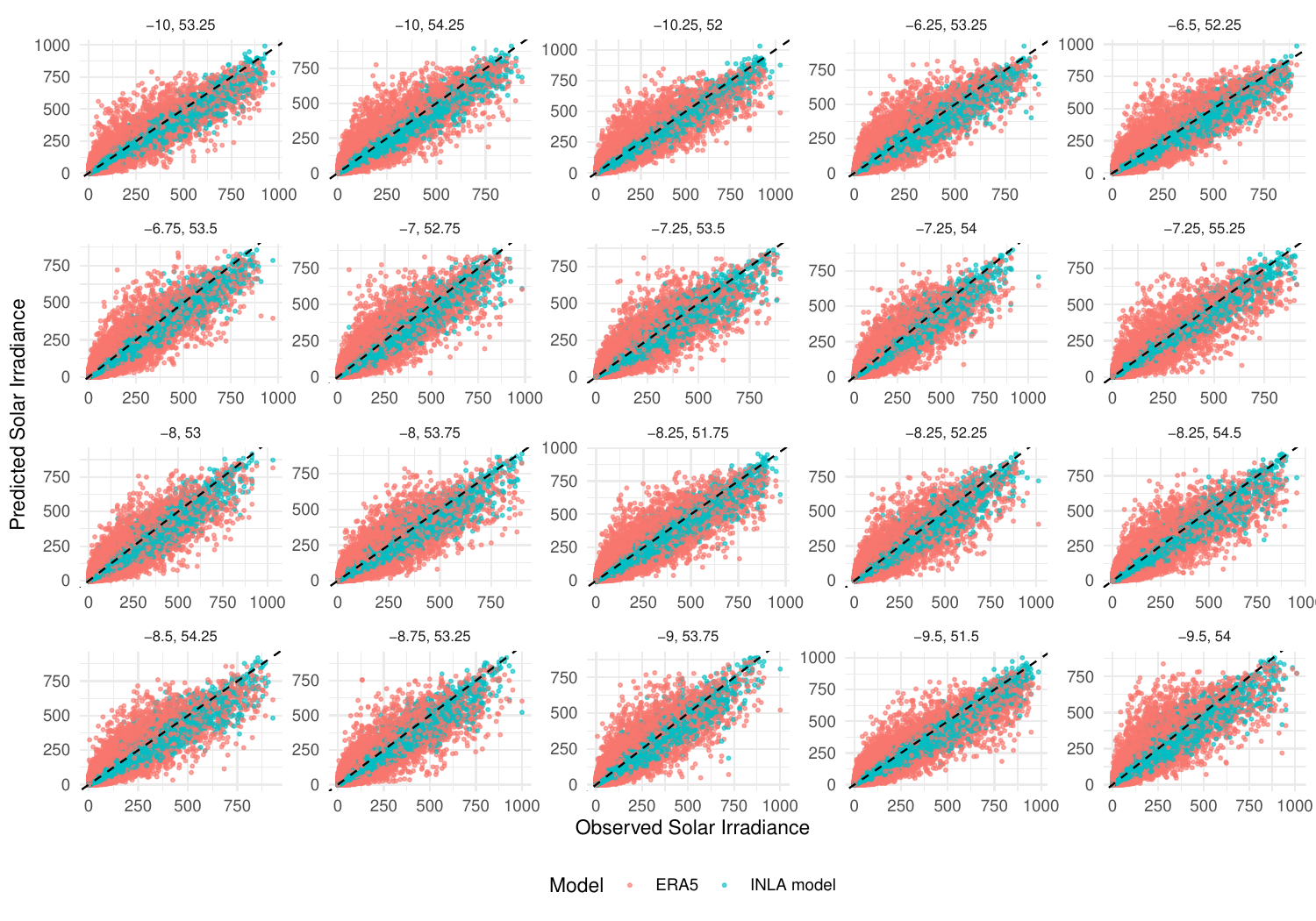}
    \caption{True versus predicted plot for INLA spatio-temporal model fit for all sites compared with ERA5 and true station for hourly resolution for full year 2024. The red represent the solar irradiance predictions using our INLA spatio-temporal model and the blue represents the solar irradiance predictions using ERA5. The true solar irradiance values are the raw solar irradiance observations from the 20 meteorological stations provided by \citet{met_eireann_web}.}
    \label{fig:true_pred_hour_era5_all_sites_2024}
\end{figure}
\end{sidewaysfigure}

\newpage

\subsubsection{10-minute resolution}
\begin{sidewaysfigure}
\begin{figure}[H]
    \centering
    \includegraphics[width=\linewidth]{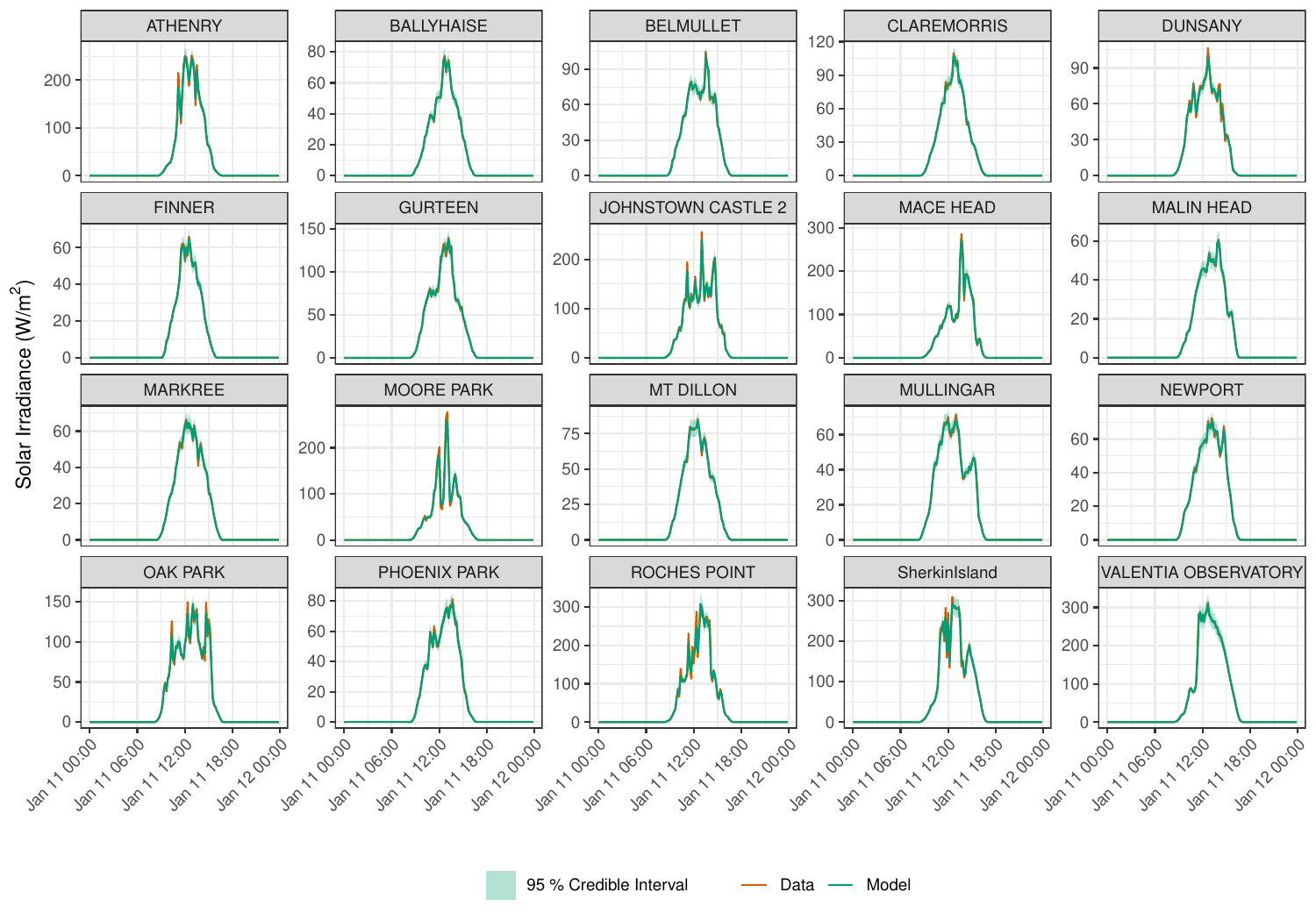}
    \caption{Model fit with 95\% credible interval for all stations for 10-minute resolution for 11th January 2024.}
    \label{fig:mod_fit_jan_10min_all_sites}
\end{figure}
\end{sidewaysfigure}

\begin{sidewaysfigure}
\begin{figure}[H]
    \centering
    \includegraphics[width=\linewidth]{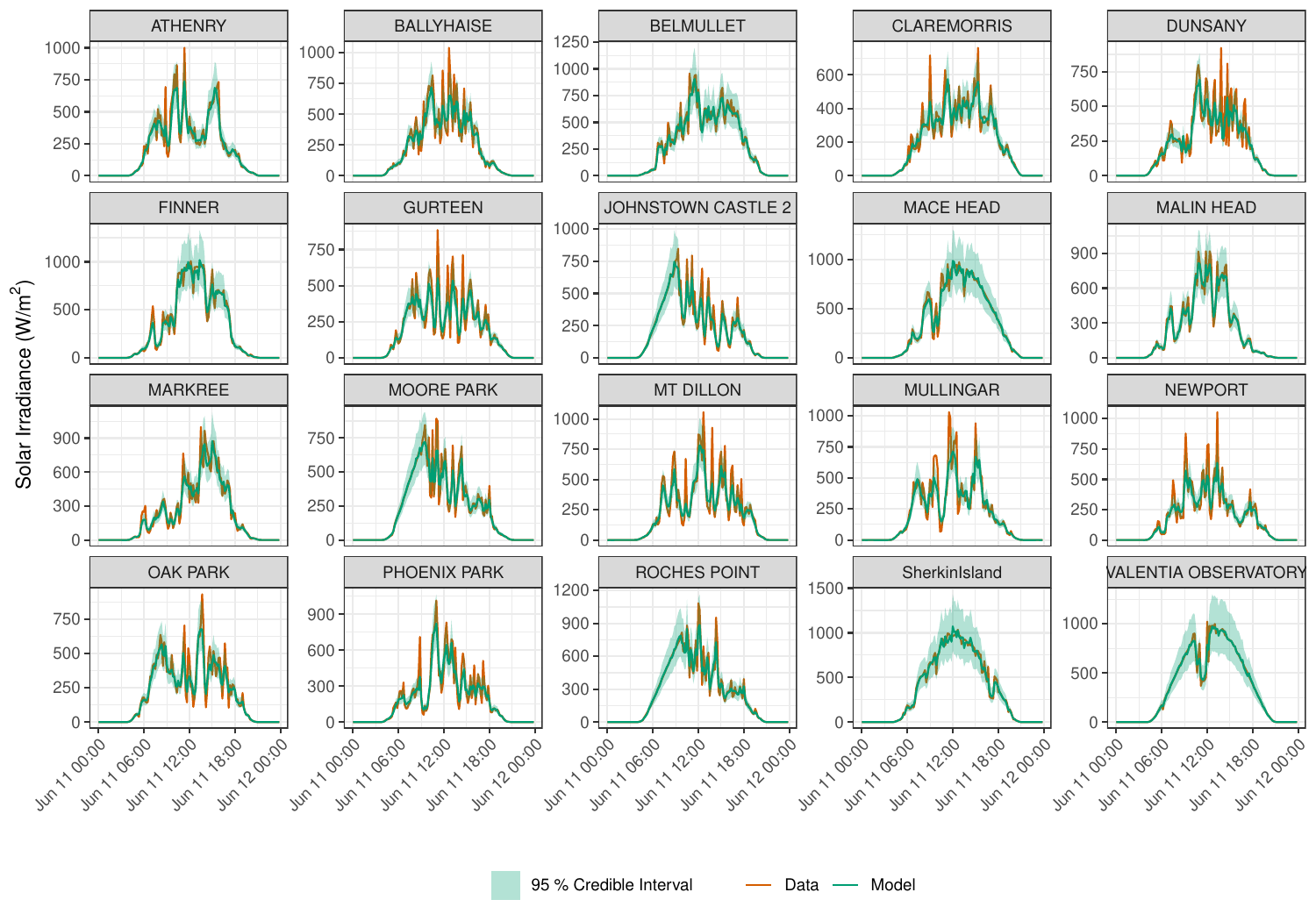}
    \caption{Model fit for with 95\% credible interval all stations for 10-minute resolution for 11th June 2024.}
    \label{fig:mod_fit_jun_10min_all_sites}
\end{figure}
\end{sidewaysfigure}

In this section, we present results from our 10-minute resolution spatio-temporal model across all 20 meteorological stations. Table \ref{tab:mod_val_full_10min} reports summary statistics for each site, while Figures \ref{fig:mod_fit_jan_10min_all_sites} and \ref{fig:mod_fit_jun_10min_all_sites} display the corresponding model fits for a representative day in January and June, respectively.
\begin{table}[ht]
\centering
\begin{tabular}{lrrrrr}
  \hline
Station & RMSE & MAE & MAPE & MBE & 95\% Coverage \\ 
  \hline
MALIN HEAD & 176.80 & 83.61 & 136.38 & -80.09 & 0.85 \\ 
  SherkinIsland & 176.61 & 78.91 & 80.68 & -73.40 & 0.96 \\ 
  VALENTIA OBSERVATORY & 167.39 & 75.78 & 100.13 & -68.98 & 0.93 \\ 
  JOHNSTOWN CASTLE 2 & 162.64 & 74.65 & 90.78 & -69.52 & 0.93 \\ 
  MACE HEAD & 157.26 & 70.19 & 80.17 & -66.37 & 0.96 \\ 
  BELMULLET & 144.20 & 63.83 & 79.97 & -57.24 & 0.95 \\ 
  ROCHES POINT & 139.65 & 60.84 & 64.27 & -53.12 & 0.96 \\ 
  FINNER & 118.03 & 49.67 & 63.20 & -40.32 & 0.96 \\ 
  BALLYHAISE & 116.42 & 49.44 & 65.48 & -40.34 & 0.97 \\ 
  MOORE PARK & 109.40 & 45.82 & 68.72 & -25.24 & 0.97 \\ 
  GURTEEN & 108.55 & 45.94 & 57.63 & -34.05 & 0.97 \\ 
  OAK PARK & 106.95 & 44.62 & 57.16 & -29.60 & 0.97 \\ 
  PHOENIX PARK & 106.45 & 44.85 & 66.04 & -31.36 & 0.95 \\ 
  ATHENRY & 100.56 & 41.54 & 65.71 & -26.39 & 0.97 \\ 
  NEWPORT & 95.38 & 39.07 & 72.41 & -14.59 & 0.95 \\ 
  MARKREE & 93.47 & 38.07 & 69.72 & -18.49 & 0.96 \\ 
  MT DILLON & 84.44 & 34.75 & 45.45 & -16.62 & 0.97 \\ 
  CLAREMORRIS & 83.94 & 35.23 & 46.59 & -14.77 & 0.96 \\ 
  DUNSANY & 83.77 & 35.22 & 40.13 & -21.67 & 0.94 \\ 
  MULLINGAR & 80.98 & 33.32 & 51.08 & -9.19 & 0.96 \\ 
   \hline
\end{tabular}
\caption{Model validation for 10-minute resolution solar irradiance using leave-one-site-out for one year.} 
\label{tab:mod_val_full_10min}
\end{table}

\newpage

\bibliography{ref.bib} 

@article{maimo2025wind,
  title={Wind and solar PV generation ramping events from farm to national level: the case of Ireland},
  author={Maim{\'o} Far, Aina and Sweeney, Conor and Flynn, Damian},
  journal={Advances in Science and Research},
  volume={22},
  pages={53--58},
  year={2025},
  publisher={Copernicus Publications G{\"o}ttingen, Germany}
}

@article{KAEWNUKULTORN2024121402,
title = {The impacts of DC/AC ratio, battery dispatch, and degradation on financial evaluation of bifacial PV+BESS systems},
journal = {Renewable Energy},
volume = {236},
pages = {121402},
year = {2024},
issn = {0960-1481},
doi = {https://doi.org/10.1016/j.renene.2024.121402},
url = {https://www.sciencedirect.com/science/article/pii/S0960148124014708},
author = {Thunchanok Kaewnukultorn and Sergio B. Sepúlveda-Mora and Steven Hegedus}
}

@book{gomez2020bayesian,
  author    = {G{\'o}mez-Rubio, Virgilio},
  title     = {Bayesian Inference with INLA},
  year      = {2020},
  publisher = {Chapman \& Hall/CRC Press},
  address   = {Boca Raton, FL}
}

@article{feng2021comparison,
  title={A comparison of zero-inflated and hurdle models for modeling zero-inflated count data},
  author={Feng, Cindy Xin},
  journal={Journal of statistical distributions and applications},
  volume={8},
  number={1},
  pages={8},
  year={2021},
  publisher={Springer}
}

@manual{SEMOpx2025,
  title        = {SEMOpx Operating Procedures, Version 9.0},
  author       = {{SEMOpx}},
  year         = {2025},
  month        = apr,
  institution  = {SEMOpx},
  note         = {Day-ahead Auctions are conducted in respect of each Trading Day, covering all Trading Periods on that day},
  url          = {https://www.sem-o.com},
}

@article{brien2017extreme,
  title={Extreme wave events in Ireland: 2012--2016},
  author={Brien, L and Renzi, Emiliano and Dudley, John Micha{\"e}l and Clancy, Colm and Dias, Fr{\'e}d{\'e}ric},
  journal={Natural Hazards and Earth System Sciences},
  pages={1--46},
  year={2017}
}

@misc{smartgriddashboard2025,
  author       = {EirGrid},
  title        = {Smart Grid Dashboard},
  howpublished = {\url{https://www.smartgriddashboard.com/roi/}},
  note         = {Last updated: 13 August 2025},
  year         = {2025},
  urldate      = {2025-09-08}
}

@inproceedings{villoz2022model,
  title={A model correcting the effect of sub-hourly irradiance fluctuations on overload clipping losses in hourly simulations},
  author={Villoz, Adrien and Wittmer, Bruno and Mermoud, Andr{\'e} and Oliosi, Michele and Bridel-Bertomeu, Agnes and Pvsyst, SA},
  booktitle={8th World Conference on Photovoltaic Energy Conversion},
  year={2022}
}

@article{fuglstad2019constructing,
  title={Constructing priors that penalize the complexity of Gaussian random fields},
  author={Fuglstad, Geir-Arne and Simpson, Daniel and Lindgren, Finn and Rue, H{\aa}vard},
  journal={Journal of the American Statistical Association},
  volume={114},
  number={525},
  pages={445--452},
  year={2019},
  publisher={Taylor \& Francis}
}

@article{organ2025enhancing,
  title={Enhancing the Accuracy of Spatio-Temporal Models for Wind Speed Prediction by Incorporating Bias-Corrected Crowdsourced Data},
  author={Organ, Eamonn and Upton, Maeve and Allard, Denis and Benoit, Lionel and Sweeney, James},
  journal={arXiv preprint arXiv:2505.24506},
  year={2025}
}

@article{https://doi.org/10.1002/met.1892,
author = {Correia, João M. and McDermott, Frank and Sweeney, Conor and Doddy, Eadaoin and Griffin, Seánie},
title = {An investigation of the regional correlation gradients between Euro-Atlantic atmospheric teleconnections and winter solar short wave radiation in northwest Europe},
journal = {Meteorological Applications},
volume = {27},
number = {2},
pages = {e1892},
doi = {https://doi.org/10.1002/met.1892},
url = {https://rmets.onlinelibrary.wiley.com/doi/abs/10.1002/met.1892},
eprint = {https://rmets.onlinelibrary.wiley.com/doi/pdf/10.1002/met.1892},
year = {2020}
}

@article{gleeson2017met,
  title={Met {\'E}ireann high resolution reanalysis for Ireland},
  author={Gleeson, Emily and Whelan, Eoin and Hanley, John},
  journal={Advances in Science and Research},
  volume={14},
  pages={49--61},
  year={2017},
  publisher={Copernicus Publications G{\"o}ttingen, Germany}
}

@Article{atmos9050163,
AUTHOR = {Nielsen, Kristian Pagh and Gleeson, Emily},
TITLE = {Using Shortwave Radiation to Evaluate the HARMONIE-AROME Weather Model},
JOURNAL = {Atmosphere},
VOLUME = {9},
YEAR = {2018},
NUMBER = {5},
ARTICLE-NUMBER = {163},
URL = {https://www.mdpi.com/2073-4433/9/5/163},
ISSN = {2073-4433},
DOI = {10.3390/atmos9050163}
}

@article{Attya25,
	author = {Attya, Mohammed and Abo-Seida, O. M. and Abdulkader, H. M. and Mohammed, Amgad M.},
	journal = {Scientific Reports},
	number = {1},
	pages = {14035},
	title = {Advanced solar radiation prediction using combined satellite imagery and tabular data processing},
	volume = {15},
	year = {2025}}

@article{BEGUERIA2026123943,
title = {A hierarchical Bayesian spatio-temporal model for estimating solar radiation from sunshine duration records},
journal = {Renewable Energy},
volume = {256},
pages = {123943},
year = {2026},
issn = {0960-1481},
doi = {https://doi.org/10.1016/j.renene.2025.123943},
url = {https://www.sciencedirect.com/science/article/pii/S0960148125016076},
author = {Santiago Beguería and Sergio M. Vicente-Serrano and José Manuel Gutiérrez-Llorente and Swen Brands and Marcos Gil-Guallar and Alejandro Royo-Aranda and María {del Mar Rondón-Velasco} and Antonio Torralba-Gallego and Yolanda Luna and Ana Morata}
}

@incollection{held2009posterior,
  title={Posterior and cross-validatory predictive checks: a comparison of MCMC and INLA},
  author={Held, Leonhard and Schr{\"o}dle, Birgit and Rue, H{\aa}vard},
  booktitle={Statistical modelling and regression structures: Festschrift in honour of ludwig fahrmeir},
  pages={91--110},
  year={2009},
  publisher={Springer}
}

@article{ridley2010modelling,
  title={Modelling of diffuse solar fraction with multiple predictors},
  author={Ridley, Barbara and Boland, John and Lauret, Philippe},
  journal={Renewable Energy},
  volume={35},
  number={2},
  pages={478--483},
  year={2010},
  publisher={Elsevier}
}

@article{wang2019echo,
  title={Echo state network based ensemble approach for wind power forecasting},
  author={Wang, Huaizhi and Lei, Zhenxing and Liu, Yang and Peng, Jianchun and Liu, Jing},
  journal={Energy Conversion and Management},
  volume={201},
  pages={112188},
  year={2019},
  publisher={Elsevier}
}

@article{dupont2022spatial,
  title={Spatial+: a novel approach to spatial confounding},
  author={Dupont, Emiko and Wood, Simon N and Augustin, Nicole H},
  journal={Biometrics},
  volume={78},
  number={4},
  pages={1279--1290},
  year={2022},
  publisher={Oxford University Press}
}

@article{chacon2024mapping,
  title={Mapping Food Insecurity in the Brazilian Amazon using a Spatial Item Factor Analysis Model},
  author={Chac{\'o}n-Montalv{\'a}n, Erick A and Parry, Luke and Giorgi, Emanuele and Torres, Patricia and Orellana, Jesem and Moraga, Paula and Taylor, Benjamin M},
  year={2024}
}

@article{brinkerink2024role,
  title={The role of spatial resolution in global electricity systems modelling},
  author={Brinkerink, Maarten and Mayfield, Erin and Deane, Paul},
  journal={Energy Strategy Reviews},
  volume={53},
  pages={101370},
  year={2024},
  publisher={Elsevier}
}

@Article{rs17060998,
AUTHOR = {Kakou, Pierre-Claver Konin and Laouali, Dungall and Aka, Boko and Osei, Janet Appiah and Ette, Nicaise Franck Kassi and Frey, Georg},
TITLE = {Multi-Timescale Validation of Satellite-Derived Global Horizontal Irradiance in Côte d’Ivoire},
JOURNAL = {Remote Sensing},
VOLUME = {17},
YEAR = {2025},
NUMBER = {6},
ARTICLE-NUMBER = {998},
URL = {https://www.mdpi.com/2072-4292/17/6/998},
ISSN = {2072-4292},
DOI = {10.3390/rs17060998}
}

@article{mathews2023systematic,
  title={Systematic bias in reanalysis-derived solar power profiles \& the potential for error propagation in long duration energy storage studies},
  author={Mathews, Duncan and Gallachoir, Brian O and Deane, Paul},
  journal={Applied Energy},
  volume={336},
  pages={120819},
  year={2023},
  publisher={Elsevier}
}

@article{BERTRAND2018306,
title = {Solar irradiation from the energy production of residential PV systems},
journal = {Renewable Energy},
volume = {125},
pages = {306-318},
year = {2018},
issn = {0960-1481},
doi = {https://doi.org/10.1016/j.renene.2018.02.036},
url = {https://www.sciencedirect.com/science/article/pii/S0960148118301812},
author = {Cédric Bertrand and Caroline Housmans and Jonathan Leloux and Michel Journée}
}

@article{gelaro2017modern,
  title={The modern-era retrospective analysis for research and applications, version 2 (MERRA-2)},
  author={Gelaro, Ronald and McCarty, Will and Su{\'a}rez, Max J and Todling, Ricardo and Molod, Andrea and Takacs, Lawrence and Randles, Cynthia A and Darmenov, Anton and Bosilovich, Michael G and Reichle, Rolf and others},
  journal={Journal of climate},
  volume={30},
  number={14},
  pages={5419--5454},
  year={2017}
}

@article{beng2004,
author = {Bengtsson, Lennart and Hagemann, Stefan and Hodges, Kevin I.},
title = {Can climate trends be calculated from reanalysis data?},
journal = {Journal of Geophysical Research: Atmospheres},
volume = {109},
number = {D11},
pages = {},
doi = {https://doi.org/10.1029/2004JD004536},
url = {https://agupubs.onlinelibrary.wiley.com/doi/abs/10.1029/2004JD004536},
eprint = {https://agupubs.onlinelibrary.wiley.com/doi/pdf/10.1029/2004JD004536},
year = {2004}
}

@Article{en12234600,
AUTHOR = {Virupaksha, Vinay and Harty, Mary and McDonnell, Kevin},
TITLE = {Microgeneration of Electricity Using a Solar Photovoltaic System in Ireland},
JOURNAL = {Energies},
VOLUME = {12},
YEAR = {2019},
NUMBER = {23},
ARTICLE-NUMBER = {4600},
URL = {https://www.mdpi.com/1996-1073/12/23/4600},
ISSN = {1996-1073},
DOI = {10.3390/en12234600}
}

@article{NUNEZMUNOZ2022122820,
title = {Development and evaluation of empirical models for the estimation of hourly horizontal diffuse solar irradiance in the United Kingdom},
journal = {Energy},
volume = {241},
pages = {122820},
year = {2022},
issn = {0360-5442},
doi = {https://doi.org/10.1016/j.energy.2021.122820},
url = {https://www.sciencedirect.com/science/article/pii/S0360544221030693},
author = {Maria {Nunez Munoz} and Erica E.F. Ballantyne and David A. Stone}
}

@article{lindgren2011explicit,
  title={An explicit link between Gaussian fields and Gaussian Markov random fields: the stochastic partial differential equation approach},
  author={Lindgren, Finn and Rue, H{\aa}vard and Lindstr{\"o}m, Johan},
  journal={Journal of the Royal Statistical Society Series B: Statistical Methodology},
  volume={73},
  number={4},
  pages={423--498},
  year={2011},
  publisher={Oxford University Press}
}

@book{williams2006gaussian,
  title={Gaussian processes for machine learning},
  author={Williams, Christopher KI and Rasmussen, Carl Edward},
  volume={2},
  year={2006},
  publisher={MIT press Cambridge, MA}
}

@article{cameletti2011comparing,
  title={Comparing spatio-temporal models for particulate matter in Piemonte},
  author={Cameletti, Michela and Ignaccolo, Rosaria and Bande, Stefano},
  journal={Environmetrics},
  volume={22},
  number={8},
  pages={985--996},
  year={2011},
  publisher={Wiley Online Library}
}

@article{bakka2018spatial,
  title={Spatial modeling with R-INLA: A review},
  author={Bakka, Haakon and Rue, H{\aa}vard and Fuglstad, Geir-Arne and Riebler, Andrea and Bolin, David and Illian, Janine and Krainski, Elias and Simpson, Daniel and Lindgren, Finn},
  journal={Wiley Interdisciplinary Reviews: Computational Statistics},
  volume={10},
  number={6},
  pages={e1443},
  year={2018},
  publisher={Wiley Online Library}
}

@article{rue2017bayesian,
  title={Bayesian computing with INLA: a review},
  author={Rue, H{\aa}vard and Riebler, Andrea and S{\o}rbye, Sigrunn H and Illian, Janine B and Simpson, Daniel P and Lindgren, Finn K},
  journal={Annual Review of Statistics and Its Application},
  volume={4},
  number={1},
  pages={395--421},
  year={2017},
  publisher={Annual Reviews}
}

@article{intro_bayes2021,
	author = {van de Schoot, Rens and Depaoli, Sarah and King, Ruth and Kramer, Bianca and M{\"a}rtens, Kaspar and Tadesse, Mahlet G. and Vannucci, Marina and Gelman, Andrew and Veen, Duco and Willemsen, Joukje and Yau, Christopher},
	date = {2021/01/14},
	doi = {10.1038/s43586-020-00001-2},
	id = {van de Schoot2021},
	isbn = {2662-8449},
	journal = {Nature Reviews Methods Primers},
	number = {1},
	pages = {1},
	title = {Bayesian statistics and modelling},
	url = {https://doi.org/10.1038/s43586-020-00001-2},
	volume = {1},
	year = {2021}}

@book{gilks1995markov,
  title={Markov chain Monte Carlo in practice},
  author={Gilks, Walter R and Richardson, Sylvia and Spiegelhalter, David},
  year={1995},
  publisher={CRC press}
}

@article{rue2009approximate,
  title={Approximate Bayesian inference for latent Gaussian models by using integrated nested Laplace approximations},
  author={Rue, H{\aa}vard and Martino, Sara and Chopin, Nicolas},
  journal={Journal of the Royal Statistical Society Series B: Statistical Methodology},
  volume={71},
  number={2},
  pages={319--392},
  year={2009},
  publisher={Oxford University Press}
}

@book{krainski2018advanced,
  title={Advanced spatial modeling with stochastic partial differential equations using R and INLA},
  author={Krainski, Elias and G{\'o}mez-Rubio, Virgilio and Bakka, Haakon and Lenzi, Amanda and Castro-Camilo, Daniela and Simpson, Daniel and Lindgren, Finn and Rue, H{\aa}vard},
  year={2018},
  publisher={Chapman and Hall/CRC}
}

@article{pfeifroth2018satellite,
  title={Satellite-based trends of solar radiation and cloud parameters in Europe},
  author={Pfeifroth, Uwe and Bojanowski, Jedrzej S and Clerbaux, Nicolas and Manara, Veronica and Sanchez-Lorenzo, Arturo and Trentmann, J{\"o}rg and Walawender, Jakub P and Hollmann, Rainer},
  journal={Advances in Science and Research},
  volume={15},
  pages={31--37},
  year={2018},
  publisher={Copernicus Publications G{\"o}ttingen, Germany}
}

@article{vzak2015cmsaf,
  title={CMSAF radiation data: New possibilities for climatological applications in the Czech Republic},
  author={{\v{Z}}{\'a}k, Michal and Mik{\v{s}}ovsk{\`y}, Ji{\v{r}}{\'\i} and Pi{\v{s}}oft, Petr},
  journal={Remote Sensing},
  volume={7},
  number={11},
  pages={14445--14457},
  year={2015},
  publisher={MDPI}
}

@misc{copernicus_ads,
  title        = "{Copernicus Atmosphere Data Store}",
  howpublished = "\url{https://atmosphere.copernicus.eu/data}",
  note         = "Accessed: 2025-06-24",
  year         = "2025",
  organization = "{Copernicus Atmosphere Monitoring Service / ECMWF}",
}

@article{solanki2013solar,
  title={Solar irradiance variability and climate},
  author={Solanki, Sami K and Krivova, Natalie A and Haigh, Joanna D},
  journal={Annual Review of Astronomy and Astrophysics},
  volume={51},
  number={1},
  pages={311--351},
  year={2013},
  publisher={Annual Reviews}
}

@misc{SARAH_website,
  doi = {10.5676/EUM_SAF_CM/SARAH/V003},
  url = "\url{https://wui.cmsaf.eu/safira/action/viewDoiDetails?acronym=SARAH_V003}",
  author = {Pfeifroth, Uwe and Kothe, Steffen and Drücke, Jaqueline and Trentmann, Jörg and Schröder, Marc and Selbach, Nathalie and Hollmann, Rainer},
  title = {Surface Radiation Data Set - Heliosat (SARAH) - Edition 3},
  publisher = {Satellite Application Facility on Climate Monitoring (CM SAF)},
  year = {2023}
}

@article{NYGARDRIISE2024112975,
title = {Benchmark of estimated solar irradiance data at high latitude locations},
journal = {Solar Energy},
volume = {282},
pages = {112975},
year = {2024},
issn = {0038-092X},
doi = {https://doi.org/10.1016/j.solener.2024.112975},
url = {https://www.sciencedirect.com/science/article/pii/S0038092X24006704},
author = {Heine {Nygard Riise} and Magnus {Moe Nygård} and Bjørn {Lupton Aarseth} and Andreas Dobler and Erik Berge}
}

@article{BRIGHT2019435,
title = {Solcast: Validation of a satellite-derived solar irradiance dataset},
journal = {Solar Energy},
volume = {189},
pages = {435-449},
year = {2019},
issn = {0038-092X},
doi = {https://doi.org/10.1016/j.solener.2019.07.086},
url = {https://www.sciencedirect.com/science/article/pii/S0038092X19307571},
author = {Jamie M. Bright}
}

@article{YANG2018876,
title = {Kriging for NSRDB PSM version 3 satellite-derived solar irradiance},
journal = {Solar Energy},
volume = {171},
pages = {876-883},
year = {2018},
issn = {0038-092X},
doi = {https://doi.org/10.1016/j.solener.2018.06.055},
url = {https://www.sciencedirect.com/science/article/pii/S0038092X18306066},
author = {Dazhi Yang}
}

@Article{su16156436,
AUTHOR = {AlFaraj, Jamal and Popovici, Emanuel and Leahy, Paul},
TITLE = {Solar Irradiance Database Comparison for PV System Design: A Case Study},
JOURNAL = {Sustainability},
VOLUME = {16},
YEAR = {2024},
NUMBER = {15},
ARTICLE-NUMBER = {6436},
URL = {https://www.mdpi.com/2071-1050/16/15/6436},
ISSN = {2071-1050},
DOI = {10.3390/su16156436}
}

@Article{sarah_satellite_paper,
AUTHOR = {Pfeifroth, U. and Dr\"ucke, J. and Kothe, S. and Trentmann, J. and Schr\"oder, M. and Hollmann, R.},
TITLE = {SARAH-3 -- satellite-based climate data records of surface solar radiation},
JOURNAL = {Earth System Science Data},
VOLUME = {16},
YEAR = {2024},
NUMBER = {11},
PAGES = {5243--5265},
URL = {https://essd.copernicus.org/articles/16/5243/2024/},
DOI = {10.5194/essd-16-5243-2024}
}

@article {MERRA2paper,
      author = "Ronald Gelaro and Will McCarty and Max J. Suárez and Ricardo Todling and Andrea Molod and Lawrence Takacs and Cynthia A. Randles and Anton Darmenov and Michael G. Bosilovich and Rolf Reichle and Krzysztof Wargan and Lawrence Coy and Richard Cullather and Clara Draper and Santha Akella and Virginie Buchard and Austin Conaty and Arlindo M. da Silva and Wei Gu and Gi-Kong Kim and Randal Koster and Robert Lucchesi and Dagmar Merkova and Jon Eric Nielsen and Gary Partyka and Steven Pawson and William Putman and Michele Rienecker and Siegfried D. Schubert and Meta Sienkiewicz and Bin Zhao",
      title = "The Modern-Era Retrospective Analysis for Research and Applications, Version 2 (MERRA-2)",
      journal = "Journal of Climate",
      year = "2017",
      publisher = "American Meteorological Society",
      address = "Boston MA, USA",
      volume = "30",
      number = "14",
      doi = "10.1175/JCLI-D-16-0758.1",
      pages=      "5419 - 5454",
      url = "https://journals.ametsoc.org/view/journals/clim/30/14/jcli-d-16-0758.1.xml"
}

@techreport{eirgrid_renewable_constraint_2024,
  author       = {{EirGrid}},
  title        = {Annual Renewable Constraint and Curtailment Report 2024},
  institution  = {EirGrid PLC},
  type         = {Technical report},
  year         = {2024},
  url          = {https://cms.eirgrid.ie/sites/default/files/publications/Annual-Renewable-Constraint-and-Curtailment-Report-2024-V1.0.pdf},
  note         = {Accessed: 2025-06-23},
}

@misc{esb_solar1gw_2024,
  author       = {{ESB Networks}},
  title        = {ESB Networks announces one Giga Watt of solar PV now energised on Ireland’s electricity network},
  howpublished = {Press release},
  institution  = {ESB Networks},
  address      = {Dublin, Ireland},
  month        = feb,
  day          = 26,
  year         = {2024},
  url          = {https://www.esb.ie/media-centre-news/press-releases/article/2024/02/26/esb-networks-announces-one-giga-watt-of-solar-pv-now-energised-on-ireland-s-electricity-network},
  note         = {Accessed: 2025-06-23}
}

@article{YANG20203,
title = {Worldwide validation of 8 satellite-derived and reanalysis solar radiation products: A preliminary evaluation and overall metrics for hourly data over 27 years},
journal = {Solar Energy},
volume = {210},
pages = {3-19},
year = {2020},
note = {Special Issue on Grid Integration},
issn = {0038-092X},
doi = {https://doi.org/10.1016/j.solener.2020.04.016},
url = {https://www.sciencedirect.com/science/article/pii/S0038092X20303893},
author = {Dazhi Yang and Jamie M. Bright}
}

@Article{geomatics1040025,
AUTHOR = {Mabasa, Brighton and Lysko, Meena D. and Moloi, Sabata J.},
TITLE = {Validating Hourly Satellite Based and Reanalysis Based Global Horizontal Irradiance Datasets over South Africa},
JOURNAL = {Geomatics},
VOLUME = {1},
YEAR = {2021},
NUMBER = {4},
PAGES = {429--449},
URL = {https://www.mdpi.com/2673-7418/1/4/25},
ISSN = {2673-7418},
DOI = {10.3390/geomatics1040025}
}

@inproceedings{davies1980calculation,
  title={Calculation of the solar radiation incident on a horizontal surface},
  author={Davies, John A and Hay, John E},
  booktitle={proc. First can. Solar radiation data workshop. Ministry of supply and services, Canada},
  pages={32--58},
  year={1980}
}

@inbook{duffie2013,
publisher = {John Wiley \& Sons, Ltd},
isbn = {9781118671603},
title = {Available Solar Radiation},
booktitle = {Solar Engineering of Thermal Processes},
chapter = {2},
pages = {43-137},
doi = {https://doi.org/10.1002/9781118671603.ch2},
url = {https://onlinelibrary.wiley.com/doi/abs/10.1002/9781118671603.ch2},
eprint = {https://onlinelibrary.wiley.com/doi/pdf/10.1002/9781118671603.ch2},
year = {2013},
author = {Duffie, J. A. and Beckman, W. A.}
}

@misc{solcast,
    year = {2024},
    author = {{Solcast}},
    url = "\url{https://solcast.com/}",
    note = "[Online; accessed 12-April-2024]"
}

@article{RIDLEY2010478,
title = {Modelling of diffuse solar fraction with multiple predictors},
journal = {Renewable Energy},
volume = {35},
number = {2},
pages = {478-483},
year = {2010},
issn = {0960-1481},
doi = {https://doi.org/10.1016/j.renene.2009.07.018},
url = {https://www.sciencedirect.com/science/article/pii/S0960148109003012},
author = {Barbara Ridley and John Boland and Philippe Lauret}
}

@article{PEREZ199789,
title = {{Comparing satellite remote sensing and ground network measurements for the production of site/time specific irradiance data}},
journal = {Solar Energy},
volume = {60},
number = {2},
pages = {89-96},
year = {1997},
issn = {0038-092X},
doi = {https://doi.org/10.1016/S0038-092X(96)00162-4},
url = {https://www.sciencedirect.com/science/article/pii/S0038092X96001624},
author = {Richard Perez and Robert Seals and Antoine Zelenka}
}

@article{kerci2024emerging,
  title={{Emerging Challenges of Integrating Solar PV in the Ireland and Northern Ireland Power Systems}},
  author={Kerci, Taulant and Hurtado, Manuel and Tweed, Simon and Escudero, Marta Val and Kennedy, Eoin and Milano, Federico},
  journal={arXiv preprint arXiv:2404.04614},
  year={2024}
}

@misc{eirgrid_system_renewable25,
  author       = {{EirGrid}},
  title        = {System and Renewable Data Reports},
  year         = {2025},
  url          = {https://www.eirgrid.ie/grid/system-and-renewable-data-reports},
  note         = {Accessed: 2025-06-23},
  institution  = {EirGrid}
}

@misc{climate_action_plan2024,
    author = {{Department of the Environment, Climate and Communications}},
    title = {{Climate Action Plan 2024}},
    url={https://www.gov.ie/en/publication/79659-climate-action-plan-2024/},
    year = {2024},
    note = "[Online; accessed 13-June-2024]"
}

@techreport{maxwell1986measuring,
  title={Measuring and modeling solar irradiance on vertical surfaces},
  author={Maxwell, Eugene L and Stoffel, Thomas L and Bird, Richard E},
  year={1986},
  institution={Solar Energy Research Inst.(SERI), Golden, CO (United States)}
}

@Article{solaR_package,
    title = {{solaR}: Solar Radiation and Photovoltaic Systems with {R}},
    author = {Oscar Perpi{\~n}{\'a}n},
    journal = {Journal of Statistical Software},
    year = {2012},
    volume = {50},
    number = {9},
    pages = {1--32},
    doi = {10.18637/jss.v050.i09}
  }

@article{doddy2021reanalysis,
  title={Which reanalysis dataset should we use for renewable energy analysis in Ireland?},
  author={Doddy Clarke, Eadaoin and Griffin, Se{\'a}nie and McDermott, Frank and Monteiro Correia, Jo{\~a}o and Sweeney, Conor},
  journal={Atmosphere},
  volume={12},
  number={5},
  pages={624},
  year={2021},
  publisher={MDPI}
}

@article{Perpinan2023,
	author = {Perpi{\~{n}}{\'{a}}n, O.},
	  title = {{Energ{\'{i}}a Solar Photovoltaic}},
	  url = {https://oscarperpinan.github.io/esf/},
	   year = {2023} }

@article{palmer2018satellite,
  title={Satellite or ground-based measurements for production of site specific hourly irradiance data: Which is most accurate and where?},
  author={Palmer, Diane and Koubli, Elena and Cole, Ian and Betts, Tom and Gottschalg, Ralph},
  journal={Solar Energy},
  volume={165},
  pages={240--255},
  year={2018},
  publisher={Elsevier}
}

@article{michalsky1988,
  title={{The Astronomical Almanac’s algorithm for approximate solar position}},
  author={Michalsky, J.},
  journal={Solar Energy},
  volume={40},
  pages={227--235},
  year={1988}
}

@article{perpinan2012solar,
  title={solaR: solar radiation and photovoltaic systems with R},
  author={Perpi{\~n}an Lamigueiro, Oscar},
  journal={Journal of Statistical Software},
  volume={50},
  number={9},
  pages={1--32},
  year={2012},
  publisher={American Statistical Association}
}

@article{ZHANG2015157,
title = {A suite of metrics for assessing the performance of solar power forecasting},
journal = {Solar Energy},
volume = {111},
pages = {157-175},
year = {2015},
issn = {0038-092X},
doi = {https://doi.org/10.1016/j.solener.2014.10.016},
url = {https://www.sciencedirect.com/science/article/pii/S0038092X14005027},
author = {Jie Zhang and Anthony Florita and Bri-Mathias Hodge and Siyuan Lu and Hendrik F. Hamann and Venkat Banunarayanan and Anna M. Brockway}
}

@article{DRIESSE2024112093,
title = {A continuous form of the Perez diffuse sky model for forward and reverse transposition},
journal = {Solar Energy},
volume = {267},
pages = {112093},
year = {2024},
issn = {0038-092X},
doi = {https://doi.org/10.1016/j.solener.2023.112093},
url = {https://www.sciencedirect.com/science/article/pii/S0038092X23007272},
author = {Anton Driesse and Adam R. Jensen and Richard Perez}
}

@misc{met_eireann_web,
    year = {2024},
    author = {{Met Éireann}},
    url = {https://www.met.ie/climate/available-data/historical-data},
    note = "[Online; accessed 10-February-2024]"
}

@misc{erigrid_soni2023,
    year = {2023},
    author = {{EirGrid and SONI}},
    title = {{Shaping our electricty future roadmap}},
    note = "[Online; accessed 17-June-2024]"
}

@misc{eirgird2022_curtail,
    author = {{EirGrid and SONI}},
    title = {{Annual Renewable Energy Constraint and Curtailment Report 2022}},
    note = "[Online; accessed 09-May-2024]",
    year = {2022}
}

@techreport{griffin2023climate,
  author       = {Griffin, Kevin and Mateus, Pedro and Lambkin, Kevin},
  title        = {Climate Data for Use in Building Design},
  institution  = {Met Éireann},
  year         = {2023},
  url          = {\url{https://www.met.ie/cms/assets/uploads/2023/03/Griffin_Mateus_Lambkin_2023_Climate-data-for-use-in-building-design.pdf}},
  note         = {Accessed: 2025-04-15}
}

@article{era5_paper,
author = {Hersbach, Hans and Bell, Bill and Berrisford, Paul and Hirahara, Shoji and Horányi, András and Muñoz-Sabater, Joaquín and Nicolas, Julien and Peubey, Carole and Radu, Raluca and Schepers, Dinand and Simmons, Adrian and Soci, Cornel and Abdalla, Saleh and Abellan, Xavier and Balsamo, Gianpaolo and Bechtold, Peter and Biavati, Gionata and Bidlot, Jean and Bonavita, Massimo and De Chiara, Giovanna and Dahlgren, Per and Dee, Dick and Diamantakis, Michail and Dragani, Rossana and Flemming, Johannes and Forbes, Richard and Fuentes, Manuel and Geer, Alan and Haimberger, Leo and Healy, Sean and Hogan, Robin J. and Hólm, Elías and Janisková, Marta and Keeley, Sarah and Laloyaux, Patrick and Lopez, Philippe and Lupu, Cristina and Radnoti, Gabor and de Rosnay, Patricia and Rozum, Iryna and Vamborg, Freja and Villaume, Sebastien and Thépaut, Jean-Noël},
title = {The ERA5 global reanalysis},
journal = {Quarterly Journal of the Royal Meteorological Society},
volume = {146},
number = {730},
pages = {1999-2049},
doi = {https://doi.org/10.1002/qj.3803},
url = {https://rmets.onlinelibrary.wiley.com/doi/abs/10.1002/qj.3803},
eprint = {https://rmets.onlinelibrary.wiley.com/doi/pdf/10.1002/qj.3803},
year = {2020}
}

@article{anderson2023pvlib,
  title={pvlib python: 2023 project update},
  author={Anderson, Kevin S and Hansen, Clifford W and Holmgren, William F and Jensen, Adam R and Mikofski, Mark A and Driesse, Anton},
  journal={Journal of Open Source Software},
  volume={8},
  number={92},
  pages={5994},
  year={2023}
}

@article{KENNY2022444,
title = {Which gridded irradiance data is best for modelling photovoltaic power production in Germany?},
journal = {Solar Energy},
volume = {232},
pages = {444-458},
year = {2022},
issn = {0038-092X},
doi = {https://doi.org/10.1016/j.solener.2021.12.044},
url = {https://www.sciencedirect.com/science/article/pii/S0038092X21010926},
author = {Darragh Kenny and Stephanie Fiedler}
}
\bibliographystyle{jmr}
\end{document}